\documentclass{article}
\usepackage{graphicx}
\usepackage{color}
\usepackage{hyperref}
\usepackage{amsmath,amssymb}
\newcommand{\be}{\begin{equation}}
\newcommand{\ee}{\end{equation}}
\newcommand{\wi}[1]{\widetilde{#1}}
\begin{document}
\vspace*{1.0cm}
\noindent
{\bf
{\Large
\begin{center}
The zig-zag road to reality
\end{center}
}
}
\vspace*{.5cm}
\begin{center}
S Colin$^{1,2,\dagger}$ and H M Wiseman$^{1,\ddagger}$
\end{center}

\begin{center}
$^1$ Centre for Quantum Dynamics, Griffith University, Brisbane, QLD 4111, Australia.
\end{center}

\begin{center}
$^2$ Perimeter Institute for Theoretical Physics, 31 Caroline Street North, ON N2L2Y5, Waterloo, Canada.
\end{center}

\begin{center}
E-mail: {$^\dagger$\texttt{s.colin@griffith.edu.au}}, {$^\ddagger$\texttt{h.wiseman@griffith.edu.au}}
\end{center}
\begin{abstract}
\noindent
In the standard model of particle physics, all fermions are fundamentally massless and only 
acquire their effective bare mass when the Higgs field condenses. Therefore, 
in a fundamental de Broglie-Bohm pilot-wave quantum field theory (valid before and after the Higgs 
condensation), position beables should be attributed to massless fermions.
In our endeavour to build a pilot-wave theory of massless fermions, which would be relevant for the study of quantum non-equilibrium in the early universe, 
we are naturally led to Weyl spinors and to particle trajectories which give meaning to the `zig-zag' picture of the electron discussed recently by Penrose. 
We show that a positive-energy massive Dirac electron of given helicity can be thought of as a superposition of positive and negative energy Weyl particles of the same helicity and that a single massive Dirac electron can in principle move luminally at all times. This is however not true for the many body situation 
required by quantum field theory and we conclude that a more natural theory arises from attributing beable status to the positions of massless Dirac particles instead of to Weyl particles.
\end{abstract}
\section{Introduction}
Inflationary cosmology will undoubtedly become a test bed for various realistic interpretations of quantum mechanics \cite{perez,valentini08}.
Indeed, even if a theory differs from standard quantum mechanics only at the very small scale, the fact 
that quantum fluctuations in the early universe (and at very small scales) act as seeds for the cosmic structure \cite{mukhanov}, can in principle render the theory testable.

Within the framework of the de Broglie-Bohm pilot-wave theory \cite{debroglie28,bohm521,bohm522}, such a possibility 
has been considered by Valentini \cite{valentini08}. One result of Valentini's analysis is that a non-equilibrium distribution for the inflaton field in the very early universe, 
although rapidly undergoing relaxation to quantum equilibrium, can leave imprints in the cosmic microwave background. 
The timescale for relaxation to quantum equilibrium is thought to be of the order of the Planck time.
Outside of the inflation paradigm, the de Broglie-Bohm theory also offers the possibility to address problems that appear to present severe conceptual difficulties 
within the standard quantum-mechanical framework \cite{peterpintoneto}.

The consideration of a massless bosonic field on expanding space (the inflaton field) fits in the program initiated by Bohm \cite{bohm522}, 
in which bosonic fields are the beables for pilot-wave quantum field theories. 
Fermionic fields, by contrast, are not beables in Bohm's program; the fermionic beables are instead typically pointlike. Several pilot-wave models for fermions have been proposed, 
for instance the Dirac-sea pilot-wave model \cite{cost07} and the Bell-type model \cite{dgtz04}.
However, all pilot-wave models for quantum field theory of fermions that have been built so far are only valid after the 
spontaneous symmetry breaking (that is, after the Higgs condenses, to use the analogy with Bose-Einstein condensates).

In the standard model of cosmology and particle physics, fermions exist before the Higgs field condenses and during that phase, all fermions (as well as gauge bosons) are massless. 
They only acquire their effective bare mass when the Higgs field condenses. 
The effective bare mass will in turn be corrected by gauge-field-mediated self-energy effects in order to yield the dressed mass. 
If relaxation to quantum equilibrium for fermions also takes place on a timescale of the order of the Planck time, it is therefore imperative to build 
a pilot-wave theory of fermions before spontaneous symmetry breaking if we want 
to study relaxation of non-equilibrium fermions.

There are several issues surrounding  the notion of  `quantum field theory before spontaneous symmetry breaking', even without considering pilot-wave theory itself.
Firstly, before the Higgs condensation, the Higgs doublet $\Phi$ is a tachyonic field. The quantization of these theories has several peculiarities \cite{feinberg}.
Secondly, all the fields should be defined on an expanding space.
Thirdly, there might be another symmetry group which is broken in $\textrm{SU}(3)\otimes\textrm{SU}(2)\otimes\textrm{U}(1)$.
Regardless of these issues, it is not disputed that all fermions are massless before the Higgs condensation.  
Therefore, there is an obvious conclusion to be drawn for the pilot-wave theory: beables should be attributed to massless fermions 
\footnote{Of course our arguments do not prove the existence of these fermionic beables in the first place and indeed it has been argued that 
one can make do purely with bosonic field beables \cite{stwe}. However bosonic field beables, apart from the Higgs field, 
are subject to the same issue: mass is not fundamental.}.
What is not clear, however, is whether this should be done using the Dirac bispinor, or the Weyl spinor. 

If we consider the Dirac equation for massless fermions and use the Weyl representation for the gamma matrices, the upper and lower 
components of the Dirac spinor become decoupled and the Dirac equation is then equivalent to a pair of Weyl equations, describing two-component 
Weyl spinors of opposite helicities. Thus the Weyl spinors seem primordial at first sight and on a first guess, 
one would attribute beables to Weyl particles in the corresponding pilot-wave theory. The true situation is not as it first appears, however. 

\begin{figure}
\begin{center}
\includegraphics[width=0.6\textwidth]{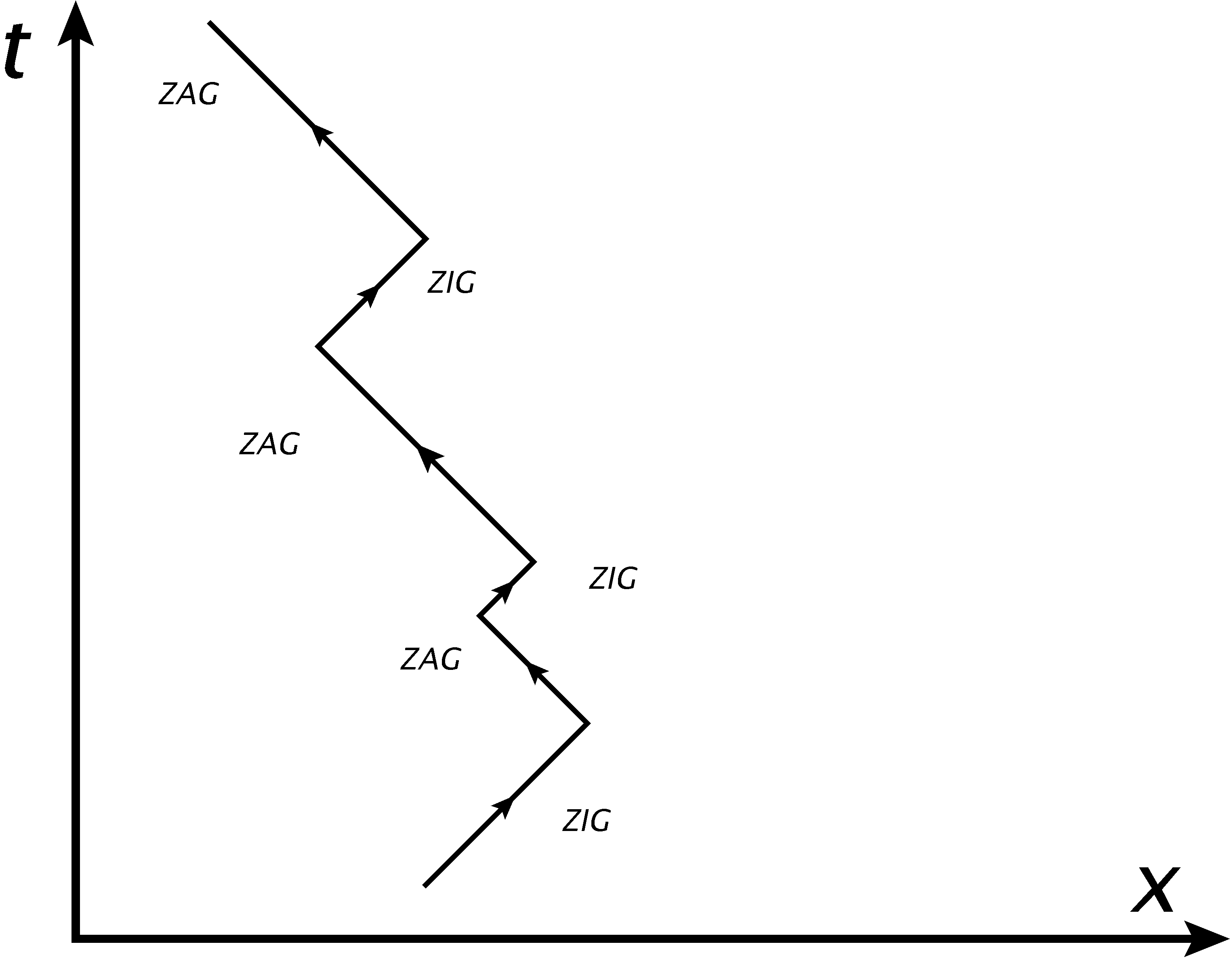}
\caption{\label{fig2}One process among an infinite number of processes contributing the propagation of the electron, in terms of zig and zag particles as formulated by 
Penrose \cite{penroseroad}.}
\end{center}
\end{figure}

Weyl spinors can also be used to describe massive Dirac fermions. In particular, 
this is how Penrose presents the standard model of particle physics in \cite{penroseroad}: the massive Dirac electron 
is represented by two types of massless Weyl fermions of opposite helicity (the `zig' and the `zag'), 
which are continually interchanged into one another. Being massless, the zig and the zag should be traveling luminally (a feature that Penrose sees as a realization of 
the `zitterbewegung'). When the zig becomes the zag, the direction of the velocity is reversed; on a coarse-grained level, jiggling between the zig and the zag motion (zigzag process) results in an apparent subluminal velocity. Each zigzag process can be illustrated by a Feynman diagram (such as shown in Fig. \ref{fig2}); however Penrose insists that 
a single zigzag process is only one element in a superposition of an infinite number of processes 
that contribute to the total electron propagator and that care must be taken in thinking about a single zigzag process as describing reality. 
Nevertheless, Penrose also asks whether one should think about the zig and the zag as describing the `real' reality, 
a question to which he suggests a positive answer (\cite{penroseroad}, p.632).

In the pilot-wave theory, there is no ambiguity about reality --- the reality is made out of the beables. 
In this article we will give a definite answer to the question raised by Penrose, by defining the pilot-wave model that corresponds to the zig-zag picture of the 
electron. In order to do that, we will first rewrite the free massive Dirac quantum field theory as an interacting Weyl quantum field theory and find the state that corresponds 
to the electron. We will show that the electron is necessarily an oscillation between two Weyl particles, but that, contrary to Penrose's picture, these Weyl particles have 
the same helicity and opposite energies. In the end, the pilot-wave model underlying the zig-zag picture is then somewhat of a hybrid between two pilot-wave models 
for quantum field theory with fermionic point-like beables: the jiggling between the zig and the zag motion 
is provided by a jump-rate (as in the stochastic model developed by D\"urr, Goldstein, Tumulka and Zangh\`{\i} \cite{dgtz04}), 
while negative energies are `real' (a feature of the deterministic Dirac-sea pilot-wave model for quantum field theory \cite{cost07}).

The zig-zag electron is a good starting point to illustrate the different ontologies that can subtend the massive Dirac 
electron and it will lead us towards the construction of a pilot-wave theory for massless fermions.
We will argue that the massless Dirac ontology is more advantageous than the Weyl ontology.
The choice of the ontology (that is, the choice of the beable) is not only a question of aesthetics. 
It  will be relevant for Valentini's conjecture \cite{valentini-phd}, according to 
which there might have been ensembles in quantum non-equilibrium in the early universe, the overwhelming majority of which 
having relaxed to quantum equilibrium.

This article is organized in the following way. In Sec. II we give a brief overview of the pilot-wave theory, including the non-relativistic case, the notion 
of quantum non-equilibrium distributions, their importance in order to test the de Broglie-Bohm pilot-wave theory  and the possible 
extensions to quantum field theory with fermionic point-like beables. In Sec. III we consider the relativistic wave equations, especially the Weyl equations 
and the corresponding pilot-wave theory. Then we will consider the case of the zig-zag electron. In Sec IV we rewrite the free massive Dirac quantum theory as an interacting 
Weyl quantum field theory, then obtain the state describing the electron in terms of Weyl particles. This allows us to construct the corresponding pilot-wave theory in Sec. V. 
Building on these results, in Sec. VI we finally address the pilot-wave theory of massless fermions. 

We use natural units in which $\hbar=c=1$ and the Minkowski metric is $g^{\mu\nu}={\rm diag}(1,-1,-1,-1)$. 
\section{The de Broglie-Bohm pilot-wave program}
The pilot-wave theory \cite{holland} was first proposed by de Broglie in 1927 \cite{debroglie28}. Bohm rediscovered it in 1952 \cite{bohm521,bohm522}, although his formulation is different (it involves accelerations, instead of velocities,  and the use of the `quantum potential'). The essential contribution of Bohm was the analysis of the measurement, the role of decoherence, and the extension to the case of the free quantized electromagnetic field. For simplicity we begin by considering the case of $N$ non-relativistic particles.

According to the de Broglie-Bohm theory, a system of $N$ spinless particles is not only described by its wave-function $\Psi(t,\vec{X})$, but also by the positions of the 
$N$ particles, that we denote by $\vec{X}(t)$, where $\vec{X}$ is a point in a configuration space $\mathbb{R}^{3n}$. As far as the laws of motion are concerned, 
$\Psi(t,\vec{X})$ always evolves according to the Schr\"odinger equation, whereas the actual configuration $\vec{X}(t)$ is guided by the wave-function, through the guidance 
equation for its velocity:
\be
\vec{V}(t)=\frac{\vec{J}(t,\vec{X})}{|\Psi(t,\vec{X})|^2}\bigg|_{\vec{X}=\vec{X}(t)}~,
\ee
Here $\vec{J}(t,\vec{X})$ is the standard quantum mechanical current, which can be defined operationally \cite{wiseman2007}.

These laws of motion have the property that if we start from an ensemble in which the beables are distributed according to $\rho(t_0,\vec{x})=|\Psi(t_0,\vec{X})|^2$ 
for some initial time $t_0$, they will be distributed according to $\rho(t,\vec{x})=|\Psi(t,\vec{X})|^2$ for any later time $t$. 
$\rho_\Psi(t,\vec{X})=|\Psi(t,\vec{X})|^2$ is referred to as the quantum equilibrium distribution \cite{valentini-phd}, or equivariant distribution \cite{durr92}.

In principle, it is also possible to consider ensembles in which the beables are distributed according to $\rho(t,\vec{X})\neq|\Psi(t,\vec{X})|^2$ and 
these would be referred to as quantum non-equilibrium distributions \cite{valentini-phd}. Such distributions may have existed in the past, and one can explain why we don't see quantum non-equilibrium today by invoking a process of relaxation to quantum equilibrium. Relaxation to quantum equilibrium is supported by numerical simulations 
\cite{valentini042,cost10}. The existence of quantum non-equilibrium in the early universe implies the possibility to test the de Broglie-Bohm theory against 
standard quantum mechanics, if this relaxation process is slow enough in certain environments.

There is no real difficulty in extending the pilot-wave approach to relativistic wave-equations, provided they can be interpreted as equations for a wave-function in the standard sense. An example of good equation is the Dirac equation. By contrast, the Klein-Gordon equation is problematic \cite{holland}.

In recent years, much progress has been made in order to extend pilot-wave ideas to quantum field theory. 
Most of these works can be seen as further developments of 
some ideas put forward by David Bohm and John Bell in 1952 and 1984 respectively. Bohm \cite{bohm522} considered the case of the free electromagnetic quantum 
field theory, and proposed to take the field itself as the beable (see \cite{struyve} for a review and discussion of the field beables approach). 
Bell \cite{bell84} considered lattice quantum field theories and proposed a stochastic model in which the beable is the collection of 
fermion-number at each point of the lattice. 
Two models have been proposed as possible extensions of the lattice stochastic Bell model to the continuum case: the stochastic Bell model with jumps \cite{dgtz04} and the 
Dirac-sea pilot-wave model for quantum field theory \cite{cost07}. Overall, the field beable approach works well for bosons whereas point-like beables can be attributed 
to fermions (where fermions means either fermions and anti-fermions of positive energy, or fermions of positive and negative energy). 

Nevertheless there are several questions that still need to be addressed. For instance, it is rarely pointed out that in pilot-wave quantum field theory 
beables must be attributed to bare particles \cite{colin_paon}. According to classical models of self-energy, the bare mass of an electron could be negative. What does it mean to attribute a position beable to a particle of negative bare mass?
In a sense, this question is alleviated in the standard model of particle physics, where all the fermions are fundamentally massless. 
Another issue is the connection between pilot-wave quantum field theories and the non-relativistic pilot wave theory (in which beables are, in effect, attributed to dressed fermions). 
\section{Relativistic wave equations for fermions}
\subsection{The Weyl equations}
The Weyl equations for massless fermions can be obtained from the standard construction that is used in order to derive the Dirac equation: 
we start from $H=\vec{\alpha}\cdot\vec{p}$, 
where $H$ is the Hamiltonian operator and $\vec{p}$ the $3$-momentum operator, impose the Hermicity of $H$ 
and the massless-particle relation $H^2=\vec{p}\cdot\vec{p}$ (hence, for plane-wave solutions, we have that $E=\pm|\vec{p}|$). 
Then the $\alpha_j$ must be Hermitian and must satisfy $\{\alpha_i,\alpha_j\}=2\delta_{ij}$. 
The simplest solutions are to take $\alpha_j=\pm\sigma_j$, where the $\sigma_j$ are the Pauli matrices:
\be
\sigma_1=\left(\begin{array}{cc} 0&1\\1&0\end{array}\right)\quad
\sigma_2=\left(\begin{array}{cc} 0&-i\\i&0\end{array}\right)\quad
\sigma_3=\left(\begin{array}{cc} 1&0\\0&-1\end{array}\right)~.
\ee
Thus we have a set of two equations:
\be
\frac{\partial}{\partial t}\psi=\mp\vec{\sigma}\cdot\vec{\nabla}\psi~.
\ee
They can be rewritten in a covariant form: $\sigma^\mu\partial_\mu\psi_R=0$ 
(resp. $\wi{\sigma}^\mu\partial_\mu\psi_L=0$), with $\sigma^\mu=(\mathbb{I}_2,\sigma_i)$ (resp. $\wi{\sigma}^\mu=(\mathbb{I}_2,-\sigma_i)$). 
The $R$ and $L$ labels, introduced to distinguish the two Weyl equations,  are related to the helicity operator, whose action on an eigenstate of 3-momentum is given by
\be 
\frac{\vec\sigma\cdot\vec{p}}{|\vec{p}|}~.
\ee
The first Weyl equation ($\sigma^\mu\partial_\mu\psi_R=0$) has right-handed positive-energy plane-wave solutions 
and left-handed negative-energy plane-wave solutions, whereas the second Weyl equation ($\wi{\sigma}^\mu\partial_\mu\psi_L=0$) 
has left-handed positive-energy plane-wave solutions and right-handed negative-energy plane-wave solutions. 
Right and left-handed eigenstates of the helicity are respectively given by:
\begin{eqnarray}
u_{R}(\vec{p})=\mathcal{N}_{\vec{p}}(|\vec{p}|\mathbb{I}_2+\vec{\sigma}\cdot\vec{p})\left(\begin{array}{cc}1 \\ 0\end{array}\right)=&\mathcal{N}_{\vec{p}}\left(\begin{array}{cc} |\vec{p}|+p_z\\p_x+ip_y\end{array}\right)\\
u_{L}(\vec{p})=\mathcal{N}_{\vec{p}}(|\vec{p}|\mathbb{I}_2-\vec{\sigma}\cdot\vec{p})\left(\begin{array}{cc}0 \\ 1\end{array}\right)=&\mathcal{N}_{\vec{p}}\left(\begin{array}{cc} -p_x+ip_y\\|\vec{p}|+p_z\end{array}\right)
\end{eqnarray}
where $\mathcal{N}_{\vec{p}}=1/\sqrt{2|\vec{p}|(|\vec{p}|+p_z)}$. 
They satisfy the following relations:
\be
u^\dagger_{R}(\vec{p})u_{R}(\vec{p})=1, u^\dagger_{L}(\vec{p})u_{R}(\vec{p})=0\textrm{ and }u^\dagger_{L}(\vec{p})u_{L}(\vec{p})=1~.~
\ee
The conserved currents are $j^\mu_R=\psi^\dagger_R\sigma^\mu\psi_R$ and  $j^\mu_L=\psi^\dagger_L\wi\sigma^\mu\psi_L$. 
We can explicitly define the eigenstates of energy-momentum:
\begin{eqnarray}
\psi_R=\begin{cases}
\frac{u_R(\vec{p})}{\sqrt{(2\pi)^3}}e^{-i|\vec{p}|t+i\vec{p}\cdot\vec{x}}\\
\frac{u_L(\vec{p})}{\sqrt{(2\pi)^3}}e^{i|\vec{p}|t+i\vec{p}\cdot\vec{x}}
\end{cases}~,\quad
\psi_L=\begin{cases}
\frac{u_L(\vec{p})}{\sqrt{(2\pi)^3}}e^{-i|\vec{p}|t+i\vec{p}\cdot\vec{x}}\\
\frac{u_R(\vec{p})}{\sqrt{(2\pi)^3}}e^{i|\vec{p}|t+i\vec{p}\cdot\vec{x}}
\end{cases}~.
\end{eqnarray}
One interesting property of the Weyl equations is that both currents are light-like, that is $g_{\mu\nu}j^\mu_{R} j^\nu_{R}=0$ and $g_{\mu\nu}j^\mu_{L} j^\nu_{L}=0$. 
It is easily shown. For instance, take
\be 
\psi_R=\left(\begin{array}{c}\psi_1 \\ \psi_2\end{array}\right)=
\left(\begin{array}{c}|\psi_1| e^{i\theta_1} \\ |\psi_2| e^{i\theta_2}\end{array}\right)~,
\ee 
then 
\be
j^\mu_R=\left(\begin{array}{c}\psi^*_1\psi_1+\psi^*_2\psi_2\\ \psi^*_1\psi_2+\psi^*_2\psi_1\\ i\psi^*_2\psi_1-i\psi^*_1\psi_2\\ \psi^*_1\psi_1-\psi^*_2\psi_2\end{array}\right)=
\left(\begin{array}{c} |\psi_1|^2+|\psi_2|^2\\ 
|\psi_1||\psi_2|\cos{(\theta_2-\theta_1)}\\ 
|\psi_1||\psi_2|\sin{(\theta_2-\theta_1)}\\ 
|\psi_1|^2-|\psi_2|^2
\end{array}\right)~,
\ee
which is light-like (the same holds for $j^\mu_L$). 

In the corresponding pilot-wave model, the Weyl fermions move with velocities 
\be
\vec{v}_R=\frac{\psi^\dagger_R\vec{\sigma}\psi_R}{\psi^\dagger_R\psi_R}\textrm{ and }\vec{v}_L=-\frac{\psi^\dagger_L\vec{\sigma}\psi_L}{\psi^\dagger_L\psi_L}~,
\ee
whose norms are $1$. This law of motion is chosen in order to ensure equivariance: if we start from an ensemble, each element being described 
by the same Weyl spinor, and if the position beables are distributed according to $j^0(t_0,\vec{x})$ over the ensemble, for some initial time $t_0$, 
they will be distributed according to 
$j^0(t,\vec{x})$ for any later time $t$. Thus Weyl fermions always move luminally. We note that this is not necessarily the case for a massless Dirac fermion, as shown in 
the next subsection.  

From the relations
\be
u^\dagger_R(\vec{p})\vec\sigma u_R(\vec{p})=\frac{\vec{p}}{|\vec{p}|}\textrm{ and }u^\dagger_L(\vec{p})\vec\sigma u_L(\vec{p})=\frac{\vec{p}}{|\vec{p}|}
\ee
we find that positive-energy plane-wave solutions move in the direction of the momentum, whereas negative-energy solutions move in the opposite direction 
of the momentum. We have something similar for a massive Dirac particle: the particle guided by a negative-energy spinor move in the opposite direction of momentum. In that case, this feature can be understood because $p^\mu=m\frac{d x^\mu}{d\tau}$ and a negative-energy particle has negative mass, hence the direction of 
velocity is opposite to that of momentum. 

\begin{figure}
\includegraphics[width=\textwidth]{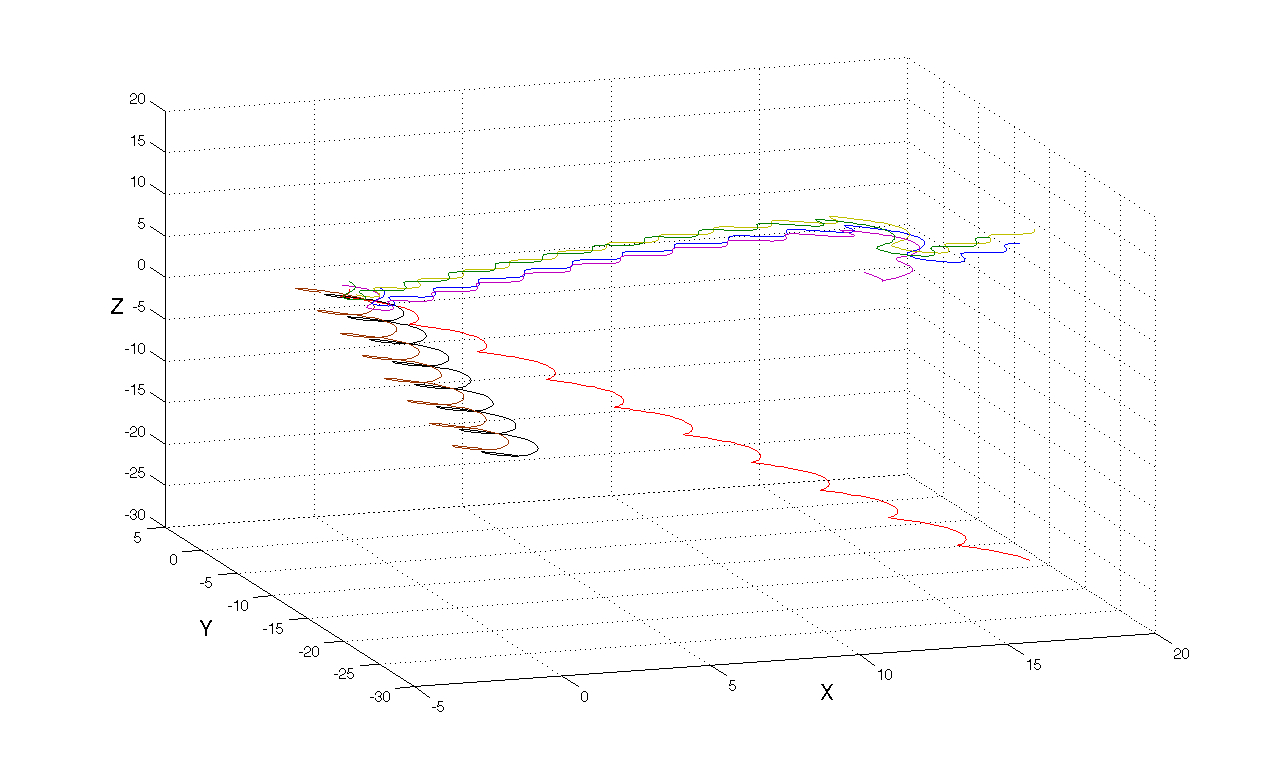}
\caption{\label{fig1}The simulations run from $t=0$ to $t=50$. 
The isolated trajectory starts from $(0,0,0)$, the couple of trajectories start from $(-1,0,0)$ and $(0,0,-1)$, while the four remaining ones 
start from $(0,0,1)$, $(0,1,0)$, $(1,0,0)$ and $(0,-1,0)$.}
\end{figure}

We plot some trajectories in Fig. \ref{fig1}. We consider a superposition of 3 Weyl spinors of positive energies and momenta $\vec{p}_1=(1,0,1)$, $\vec{p}_2=(-1,-2,-1)$ and $\vec{p}_3=(1,-1,1)$. Each spinor has equal weight and the phases for 2 and 3 are $e^{i 4}$ and $e^{i 9}$.

Finally, we say a few words about relaxation to quantum equilibrium. This is measured by the coarse-grained H-function
\be
\bar{H}(t)=\int d^3x \bar{\rho} \ln(\bar{\rho}/\overline{|\psi|^2})~.
\ee 
In earlier works \cite{valentini-phd,valentini042}, the relaxation time was defined through the second-order derivative of $\bar{H}(t)$ at $t=t_0$. 
However, as discussed in \cite{toruva}, this timescale can only characterize the relaxation close to $t_0$, 
where $\bar{H}(t)$ is not a decaying exponential (since $\dot{\bar{H}}(t)\big|_{t=t_0}=0$). 
A new estimate for the `true' relaxation time (once $\bar{H}(t)$ undergoes exponential decay), for the case of a scalar particle in a square box of side-length $L$, was given in \cite{toruva} 
and it is in agreement with the numerical simulations performed in \cite{toruva}. 
This new estimate of the relaxation time is obtained by assuming that the displacement of the particle during the relaxation time should be of the order of $L$, 
where $L$ is roughly the displacement needed in order to hit one of the nodes, from which vorticity, chaos and relaxation originate.

Let us first consider an ensemble of out-of-equilibrium Weyl fermions which are not confined to the interior of a box. 
From purely dimensional arguments, one expects the relaxation time, if any,  to be given by 
\be
\frac{\lambda}{c}\left(\frac{c\hbar}{\Delta E\lambda}\right)^{a}~,
\ee
where $\lambda$ is the coarse-graining length, $\Delta E$ the energy spreading and $a$ a positive number.

Turning now to Weyl fermions inside a box, one might be tempted to repeat the derivation of the relaxation 
timescale given in \cite{toruva}. However such an analysis can't be repeated. 
The reason is that Weyl spinors, like any spinor, typically do not have nodes\footnote{For instance, for a Weyl spinor $\left(\begin{array}{c}\psi_1\\\psi_2\end{array}\right)$, 4 conditions need to be satisfied in order to get a node: $\mathfrak{Re}(\psi_1)=0$, $\mathfrak{Im}(\psi_1)=0$, 
$\mathfrak{Re}(\psi_2)=0$ and $\mathfrak{Im}(\psi_2)=0$. If there are 3 spatial dimensions, each of these conditions define a surface. 
Typically four different surfaces do not intersect and therefore we typically don't expect any node. Given that the nodes are the origin of relaxation 
to quantum equilibrium for scalar particles, the fact that nodes are untypical does not mean that out-of-equilibrium Weyl fermions 
will never reach equilibrium. Indeed the important feature of the nodes is the associated vorticity (\cite{valentini042,cost10,efthymiopoulos}). 
For scalar particles, there is only vorticity in the vicinity of a node but for a spinor particle, vorticity is not necessarily associated to a node.}. 
Therefore we can't use the argument that after a typical displacement $L$, the particle is going to hit a node.

\subsection{The Dirac equation}
Both Weyl equations can be rewritten as a single equation obeyed by a 4-component Dirac spinor. If we use the Weyl representation of the Dirac algebra, given by 
\be\label{weylgamma}
\gamma^0=\left(\begin{array}{cc} 0&1\\1&0\end{array}\right),\quad
\gamma^i=\left(\begin{array}{cc} 0&\sigma_i\\-\sigma_i&0\end{array}\right)~,\quad
\gamma^\mu=\left(\begin{array}{cc} 0&\sigma^\mu\\\wi{\sigma}^\mu&0\end{array}\right).
\ee
then $\gamma^\mu\partial_\mu\psi=0$, with $\psi=\left(\begin{array}{c}\psi_L \\ \psi_R\end{array}\right)$, reduces to
\begin{eqnarray}
\partial_t\psi_R=-\vec{\sigma}\cdot\vec{\nabla}\psi_R\\
\partial_t\psi_L=\vec{\sigma}\cdot\vec{\nabla}\psi_L~.
\end{eqnarray}

The massless case just described generalizes for massive particles as the Dirac equation:
\be
(i\gamma^\mu\partial_\mu-m)\psi(t,\vec{x})=0~,
\ee
where $\psi(t,\vec{x})$ is a 4-spinor and $\{\gamma^\mu,\gamma^\nu\}=2 g^{\mu\nu}$. It has a conserved 4-current 
with positive 0-component:
\be
j^\mu=\bar{\psi}\gamma^\mu\psi=\psi^\dagger_R\sigma^\mu\psi_R+\psi^\dagger_L\wi\sigma^\mu\psi_L
\ee
Note that the current, being the sum of two light-like currents, is not necessarily light-like itself.
The corresponding pilot-wave theory follows by attributing a position $\vec{x}(t)$ to the Dirac electron and defining its velocity as 
\be
\vec{v}(t)=\frac{\psi^\dagger_R\vec\sigma\psi_R-\psi^\dagger_L\vec\sigma\psi_L}{\psi^\dagger_R\psi_R+\psi^\dagger_L\psi_L}\bigg|_{\vec{x}=\vec{x}(t)}~.
\ee

As an aside remark, in the pilot-wave theory for the Dirac electron, the 
zittebewegung can be seen from the electron trajectory \cite{holland}. A recent ion-trap experiment (analog to a $1+1$ Dirac equation) also 
supports the reality of the zitterbewegung \cite{nature}. Tausk and Tumulka have also asked the question whether the Dirac electron could move luminally \cite{tatu}.
\subsection{The zig-zag picture of the electron}
Let us consider the massive Dirac equation in the Weyl representation for the $\gamma$-matrices:
\begin{eqnarray}
i\sigma^\mu\partial_\mu\psi_R=m\psi_L,\\
i\widetilde{\sigma}^\mu\partial_\mu\psi_L=m\psi_R ~.
\end{eqnarray}
The massive Dirac equation is then equivalent to a pair of Weyl equations, where each Weyl spinor acts as a source for its partner. Hence a massive 
Dirac electron can be thought of as being composed of two Weyl spinors. 

According to Penrose, the Dirac electron can be thought of as an oscillation between two particles of opposite helicities (\cite{penroseroad}, p.630). 
Penrose refers to the Weyl particle of negative (resp. positive) helicity as the zig (resp. zag) particle. 
Being massless, they move at the speed of light and, still according to Penrose, this is a realization of the 
zitterbewegung, which says that the instantaneous velocity of an electron is always measured to be that of light. 

Because $\psi_R$ (resp. $\psi_L$) describes positive-energy right-handed (resp. left-handed) solutions and negative-energy left-handed 
(resp. right-handed) solutions, a positive-energy right-handed Dirac electron, made of two Weyl spinors $\psi_R$ and $\psi_L$, can be one of the following two: 
\begin{itemize}
\item a superposition of positive-energy right-handed and left-handed Weyl spinors, as advocated by Penrose,
\item a superposition of positive and negative energy right-handed Weyl spinors.
\end{itemize}

In the following, we will show that the second option is the correct one in pilot-wave theory. That is, picturing a positive-energy massive Dirac electron of given helicity as an oscillation between two Weyl particles only makes sense if the Weyl particles have the same helicity but positive and negative energies.
\footnote{If, instead of helicity, we consider chirality, defined by the operator $\gamma^5=i\gamma^0\gamma^1\gamma^2\gamma^3$, the second option amounts to saying that 
a positive-energy right-handed electron is an oscillation between a positive-energy Weyl particle of positive chirality and a negative-energy Weyl particle of negative chirality. 
Indeed, in the case of massless particles, although chirality and helicity are equivalent for positive-energy solutions, they are opposite for negative-energy solutions. 
This would provide a way to understand Penrose's claim (by replacing helicity by chirality each time it is mentioned). As far as the forthcoming analysis is concerned, 
this distinction between chirality and helicity is not important.} 
In order to show that, we shall rewrite the free Dirac massive Dirac quantum field theory as a Weyl quantum field theory with interactions.
\section{The zig-zag electron in standard QFT}
\subsection{Free massless Dirac QFT = free Weyl(s) QFT}
\begin{figure}
\begin{center}
\includegraphics[width=0.4\textwidth]{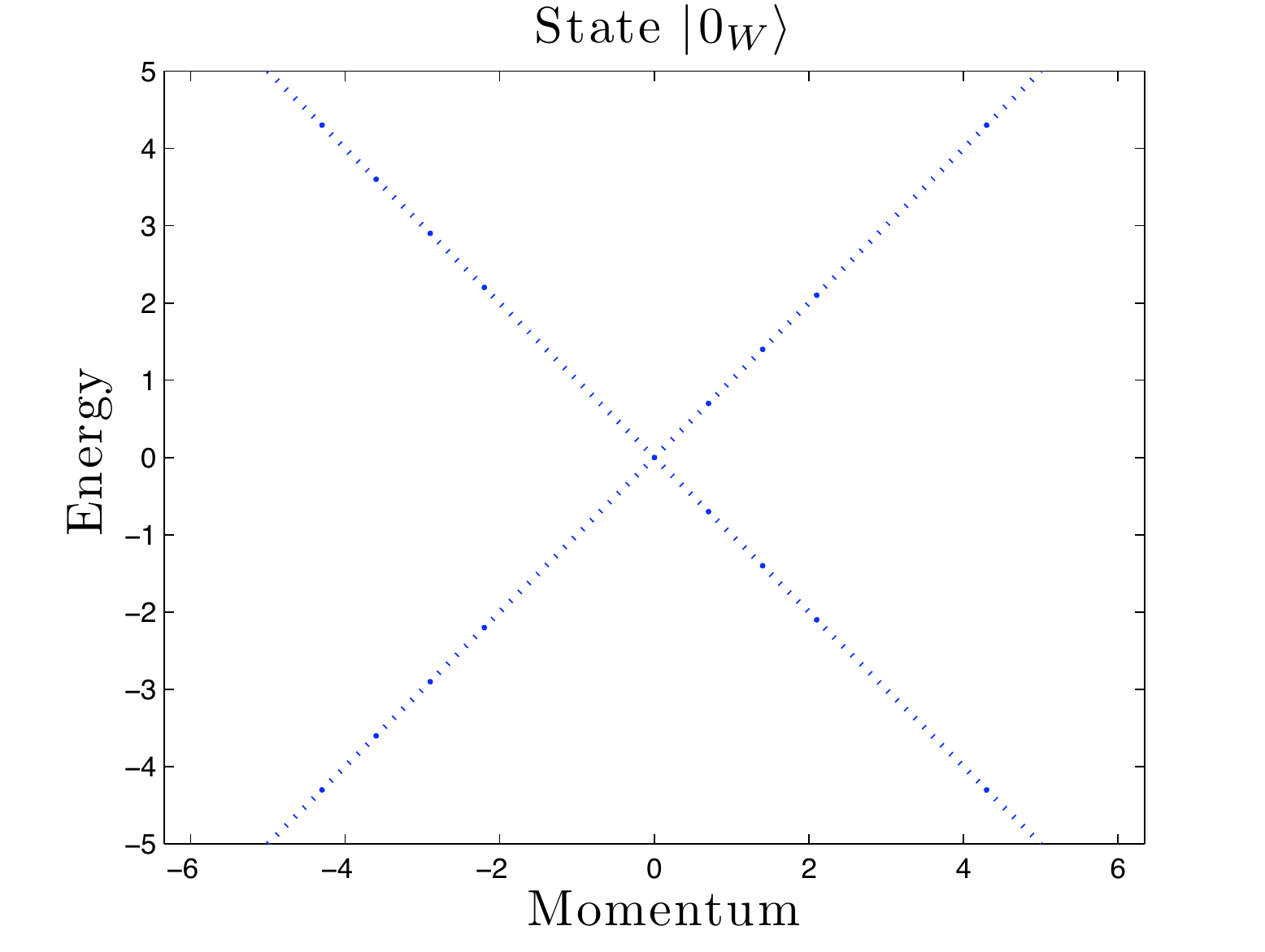}
\includegraphics[width=0.4\textwidth]{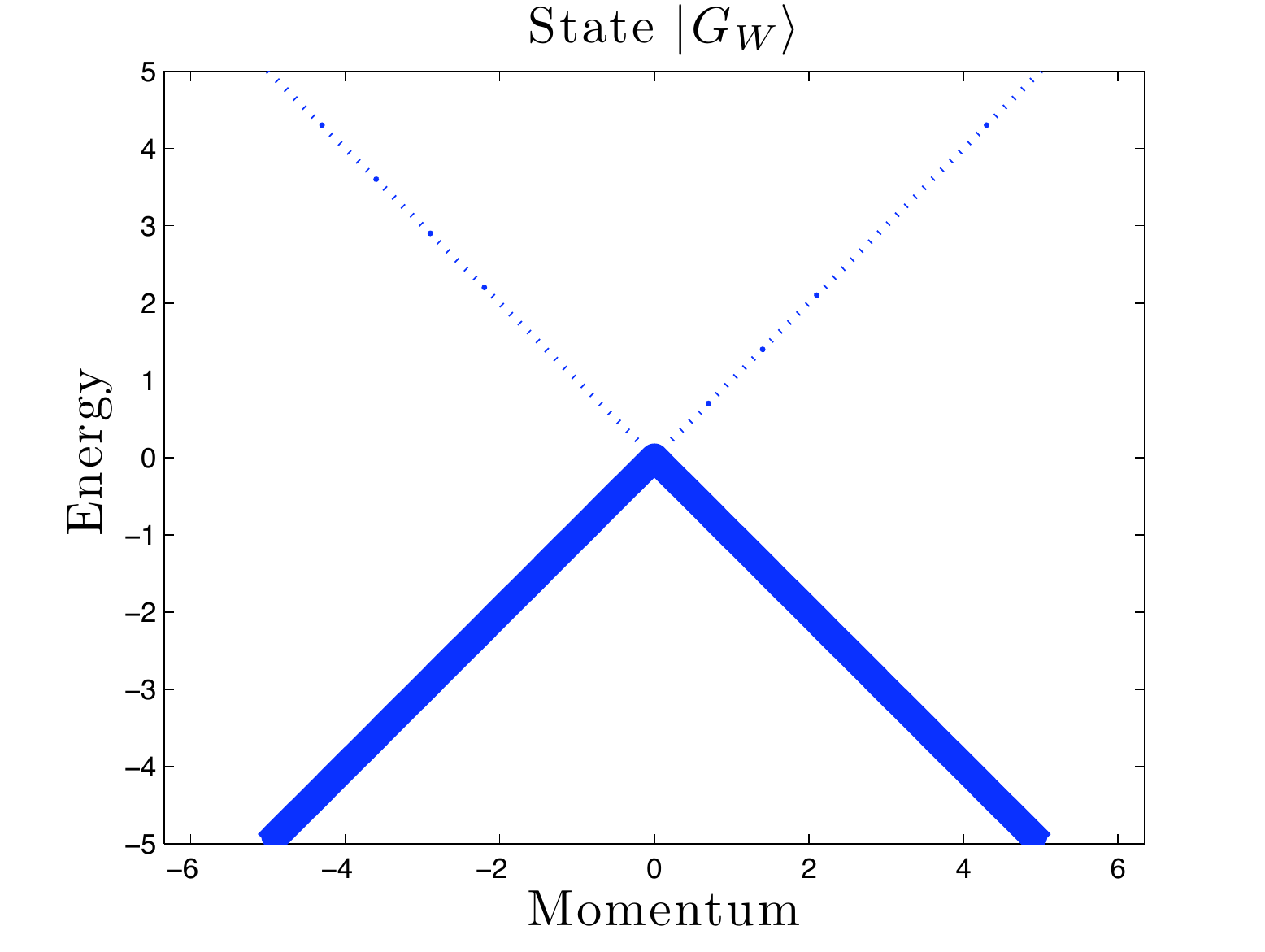}
\end{center}
\caption{\label{fig3}The states $|0_W\rangle$ and $|G_W\rangle$ in the free Weyl(s) QFT. 
Bold (dotted) lines correspond to (un)occupied levels.}
\end{figure}

We start from the free Dirac Lagrangian for massless fermions:
\be\label{masslessdiraclag}
\mathcal{L}_0=i\bar{\psi}\gamma^\mu\partial_\mu\psi=i\psi^\dagger\gamma^0\gamma^\mu\partial_\mu\psi~.
\ee
If we use the Weyl representation for the $\gamma$-matrices (Eq. (\ref{weylgamma})), we get:
\begin{eqnarray}
\mathcal{L}_0=i\psi^\dagger_{L}\wi{\sigma}^\mu\partial_\mu\psi_{L}+i\psi^\dagger_{R}\sigma^\mu\partial_\mu\psi_{R}\nonumber\\
=i\psi^\dagger_L(\partial_t-\vec{\sigma}\cdot\vec{\nabla})\psi_L+i\psi^\dagger_R(\partial_t+\vec{\sigma}\cdot\vec{\nabla})\psi_R~.~
\end{eqnarray}
Then the free Hamiltonian reads:
\be
H_0=\int d^3x [i\psi^\dagger_L(t,\vec{x})\vec{\sigma}\cdot\vec{\nabla}\psi_L(t,\vec{x})-i\psi^\dagger_R(t,\vec{x})\vec{\sigma}\cdot\vec{\nabla}\psi_R(t,\vec{x})]~.
\ee
After canonical quantization, the expressions for the fields are:
\begin{align}\label{wfields}
%\fl
\hat\psi_L(t,\vec{x})=\frac{1}{\sqrt{(2\pi)^3}}\int d^3p~[u_L(\vec{p})\hat{c}_L(\vec{p})e^{-i|\vec{p}|t+i\vec{p}\cdot\vec{x}}
+u_R(\vec{p})\hat\zeta_R(\vec{p})e^{i|\vec{p}|t+i\vec{p}\cdot\vec{x}}]\quad{\rm and }
\nonumber\\
%\fl
\hat\psi_R(t,\vec{x})=\frac{1}{\sqrt{(2\pi)^3}}\int d^3p~[u_R(\vec{p})\hat{c}_R(\vec{p})e^{-i|\vec{p}|t+i\vec{p}\cdot\vec{x}}
+u_L(\vec{p})\hat\zeta_L(\vec{p})e^{i|\vec{p}|t+i\vec{p}\cdot\vec{x}}]~.~
\end{align}
The canonical anti-commutation relations are given by
\begin{eqnarray}
\{\hat\psi_{\chi,a}(\vec{x}),\hat\psi^\dagger_{\chi,a}(\vec{y})\}=\delta^3(\vec{x}-\vec{y}),\nonumber\\
\{\hat{c}_{\chi}(\vec{p}),\hat{c}^\dagger_{\chi}(\vec{q})\}=\delta^3(\vec{p}-\vec{q})\textrm{ and }\nonumber\\
\{\hat\zeta_{\chi}(\vec{p}),\hat\zeta^\dagger_{\chi}(\vec{q})\}=\delta^3(\vec{p}-\vec{q})~.
\end{eqnarray}
with $\chi\in\{L,R\}$ and where all the remaining anti-commutators vanish. 
If we substitute the previous expressions in the Hamiltonian, we find:
\be
\hat{H}_0=\sum_{\chi\in\{L,R\}}\int d^3p ~|\vec{p}|[\hat{c}^\dagger_\chi(\vec{p})\hat{c}_\chi(\vec{p})-\hat\zeta^\dagger_\chi(\vec{p})\hat\zeta_\chi(\vec{p})]~.
\ee
Hence the operator $\hat{c}^\dagger_\chi(\vec{p})$ creates a positive-energy Weyl particle of 3-momentum $\vec{p}$ and helicity $\chi$ while the operator
$\hat\zeta^\dagger_\chi(\vec{p})$ creates a negative-energy Weyl particle of 3-momentum $\vec{p}$ and helicity $\chi$.
We define $|0_W\rangle$ as the state that does not contain any positive or negative energy Weyl particle:  
\be
\hat{c}_\chi(\vec{p})|0_W\rangle=0\quad \hat\zeta_\chi(\vec{p})|0_W\rangle=0\quad\forall\vec{p},\chi~.
\ee
Note that $|0_W\rangle$ is not the lowest state of energy for $\widehat{H}_0$. For the lowest state of energy (the ground state), we 
use the notation $|G_W\rangle$ (it is the state in which all negative-energy levels are filled).
If we reinterpret the annihilation of a negative-energy particle as the creation of an anti-particle
\be
\zeta_\chi(\vec{p})=d^\dagger_\chi(-\vec{p})~,
\ee
then the state $|G_W\rangle$ is defined by 
\be
\hat{c}_\chi(\vec{p})|G_W\rangle=0\quad \hat{d}_\chi(\vec{p})|G_W\rangle=0\quad\forall\vec{p},\chi~.
\ee
Both states are represented in Fig. \ref{fig3}.
\subsection{Free massive Dirac QFT = Weyl(s) QFT with interaction}
Now we consider the addition of the mass term $-m\int d^3x \bar{\psi}\psi=-m\int d^3x (\psi^\dagger_L\psi_R+\psi^\dagger_R\psi_L)$ to the 
free Dirac Lagrangian (Eq. (\ref{masslessdiraclag})), in order to obtain the  massive Dirac Lagrangian. This term will give rise to the interaction Hamiltonian  
\be
\hat{H}_I=m\sum_{\chi\in\{L,R\}}\int d^3p [\hat{c}^\dagger_\chi(\vec{p})\hat\zeta_\chi(\vec{p})+\hat\zeta^\dagger_\chi(\vec{p})\hat{c}_\chi(\vec{p})]~.
\ee
The total Hamiltonian is then
\begin{align}\label{Hzeta}
%\fl
\hat{H}=\sum_{\chi\in\{L,R\}}\int d^3p[|\vec{p}|\hat{c}^\dagger_\chi(\vec{p})\hat{c}_\chi(\vec{p})-|\vec{p}|\hat\zeta^\dagger_\chi(\vec{p})\hat\zeta_\chi(\vec{p})
+m\hat{c}^\dagger_\chi(\vec{p})\hat\zeta_\chi(\vec{p})+m\hat\zeta^\dagger_\chi(\vec{p})\hat{c}_\chi(\vec{p})]~.~
\end{align}
In the particle-antiparticle picture, after normal-ordering, the expression of the total Hamiltonian is:
\begin{eqnarray}\label{Hd}
:\hat{H}:=\sum_{\chi\in\{L,R\}}\int d^3p [|\vec{p}|\hat{c}^\dagger_\chi(\vec{p})\hat{c}_\chi(\vec{p})+|\vec{p}|\hat{d}^\dagger_\chi(\vec{p})\hat{d}_\chi(\vec{p})\nonumber\\
+m\hat{c}^\dagger_\chi(\vec{p})\hat{d}^\dagger_\chi(-\vec{p})+m\hat{d}_\chi(-\vec{p})\hat{c}_\chi(\vec{p})]~.\quad
\end{eqnarray}
For further comparison, we also introduce the states corresponding to $|0_W\rangle$ and $|G_W\rangle$:
we call them $|0_D\rangle$ and $|G_D\rangle$. $|0_D\rangle$ is the state that doesn't contain any positive or negative energy massive 
Dirac electron, whereas $|G_D\rangle$ is the state of lowest energy for the massive Dirac QFT (in which all negative energy states are filled).
\begin{figure}
\begin{center}
\includegraphics[width=0.4\textwidth]{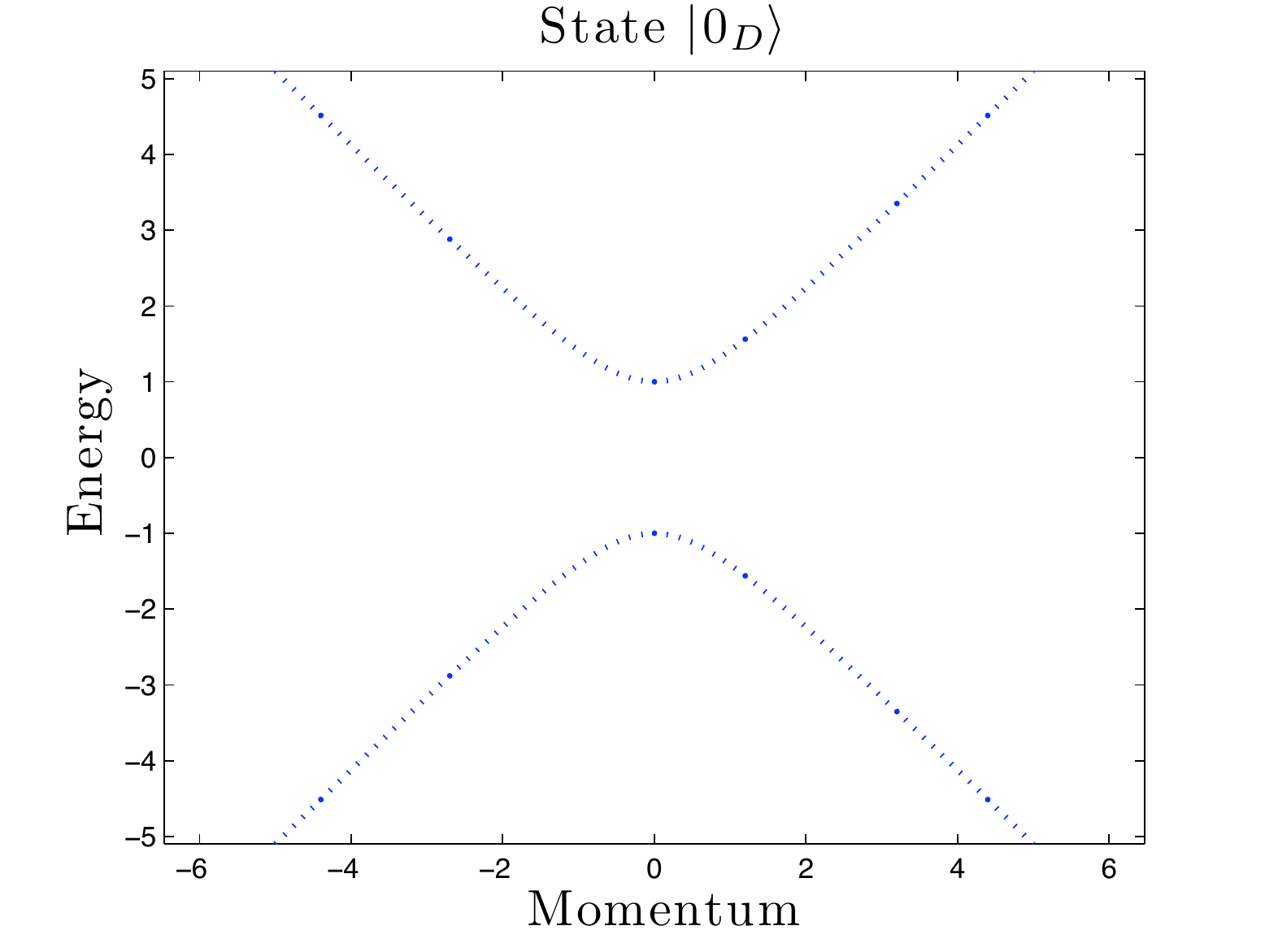}
\includegraphics[width=0.4\textwidth]{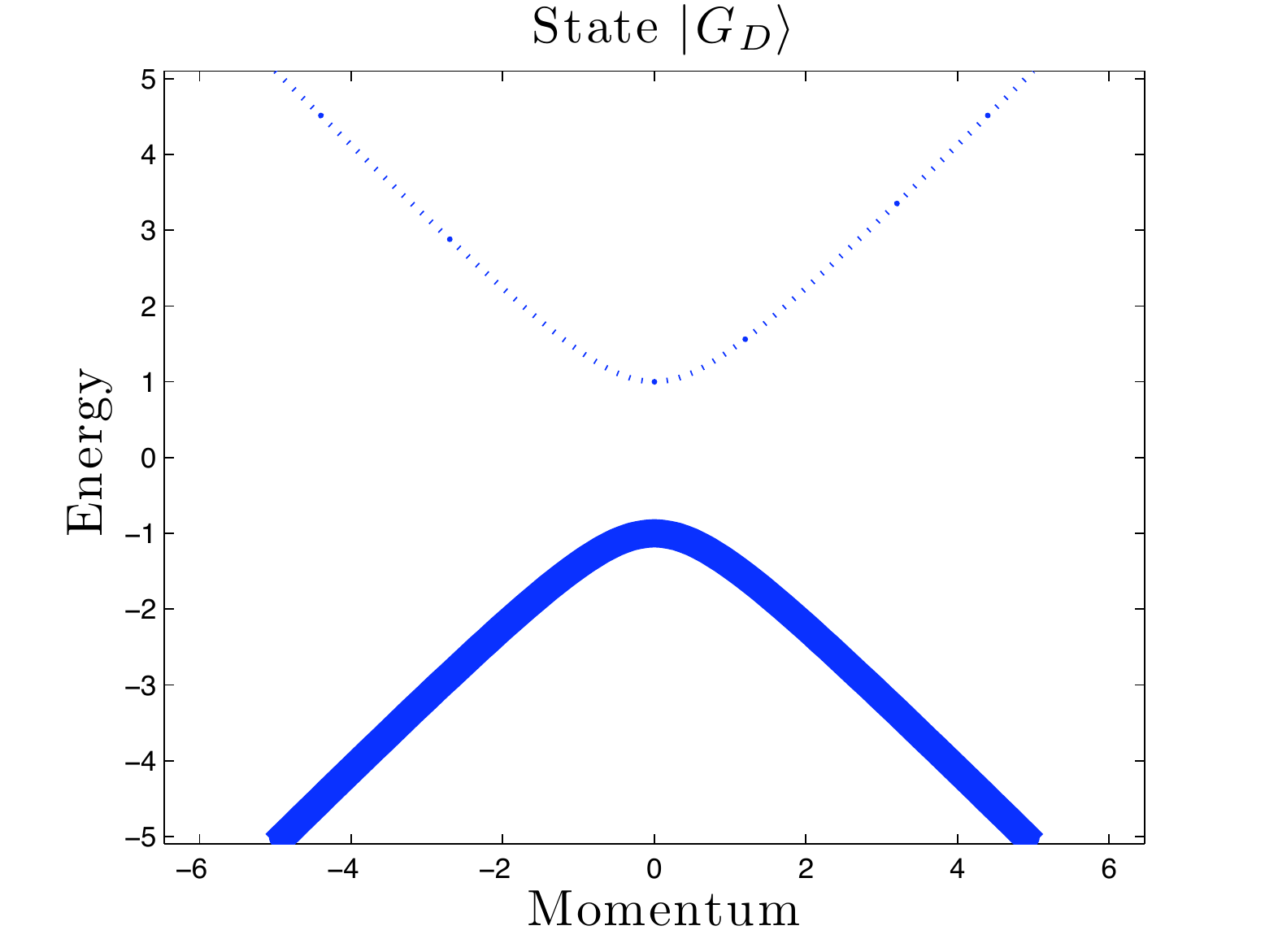}
\end{center}
\caption{\label{fig4}The states $|0_D\rangle$ and $|G_D\rangle$ in the free massive Dirac QFT. 
For this figure and the next ones in this section, we assume that the mass of the Dirac particle is equal to $1$.}
\end{figure}
$|0_D\rangle$ and $|G_D\rangle$ are represented in Fig. \ref{fig4}.  We also have that 
\be
|0_D\rangle=|0_W\rangle~.
\ee
In terms of Weyl particles and anti-particles, $|G_D\rangle$ is a complex state: it is a superposition of the state
containing no Weyl particle or anti-particle ($|G_W\rangle$), plus states containing one pair of Weyl particle-antiparticle, plus states containing two pairs of 
Weyl particle-antiparticle etc. This is illustrated in Fig. (\ref{fig7a}).
\begin{figure}
\begin{center}
\includegraphics[width=0.8\textwidth]{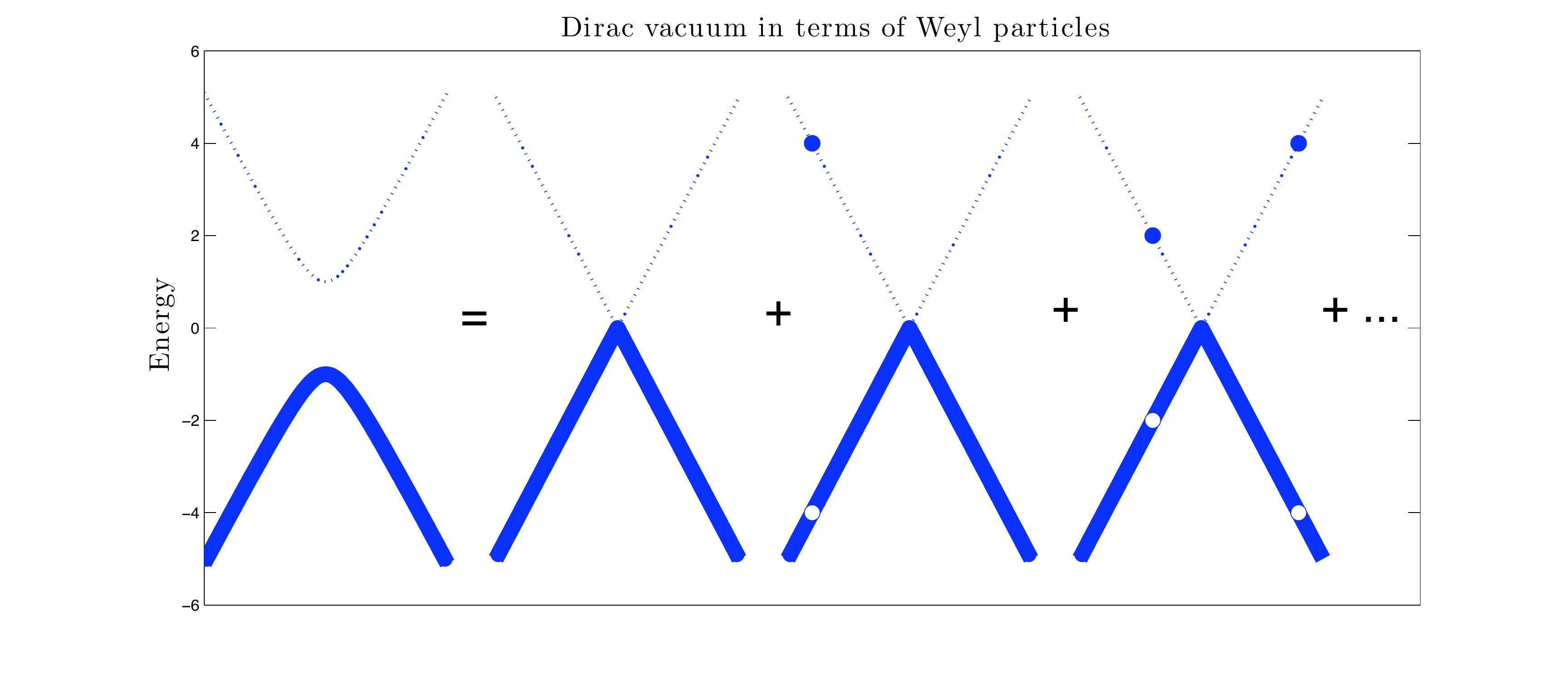}
\end{center}
\caption{\label{fig7a}The Dirac vacuum in terms of Weyl particles and anti-particles. The white dots represent holes in the sea.}
\end{figure}
\subsection{The zig-zag electron}
\begin{figure}
\begin{center}
\includegraphics[width=0.4\textwidth]{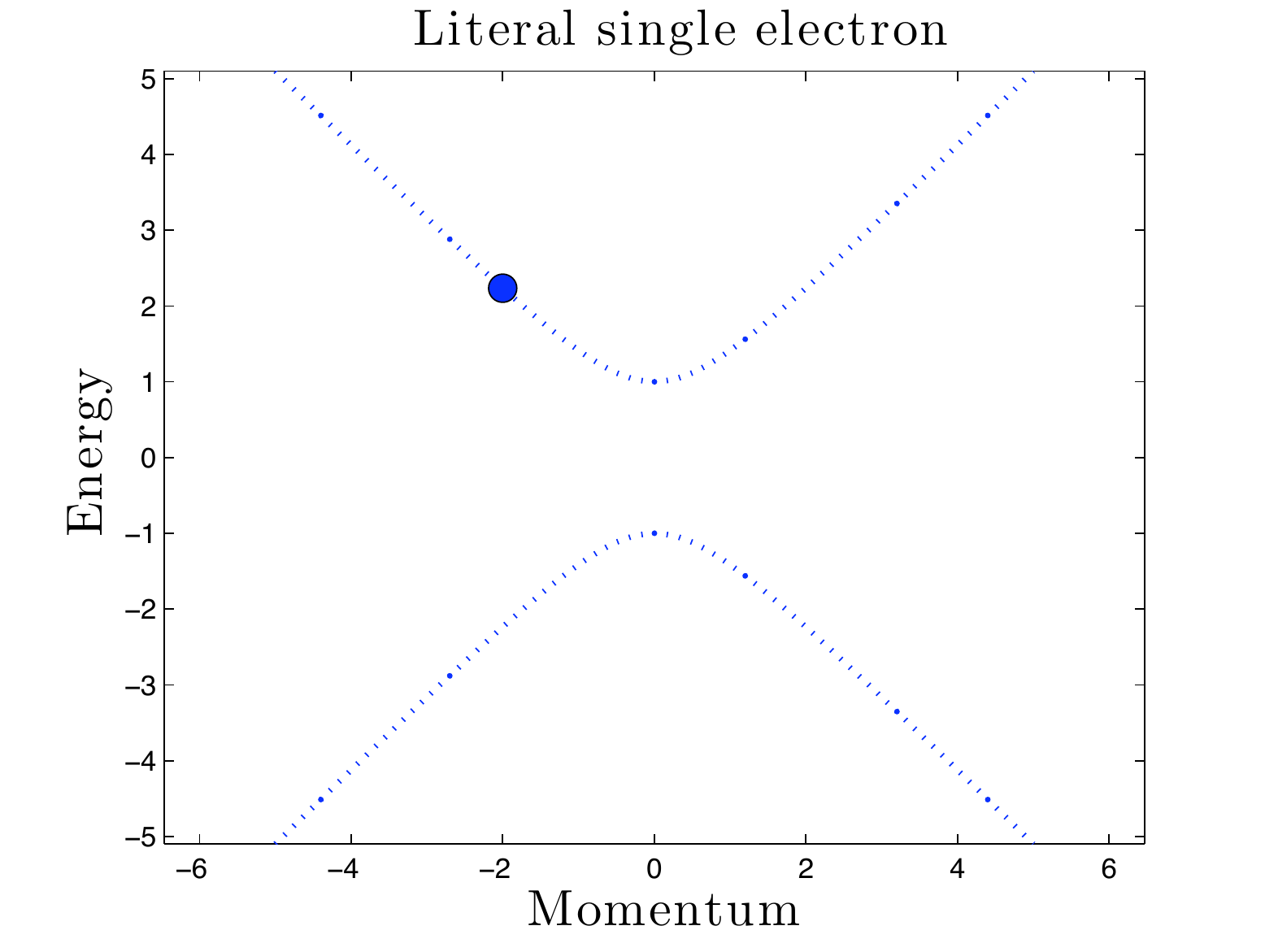}
\includegraphics[width=0.4\textwidth]{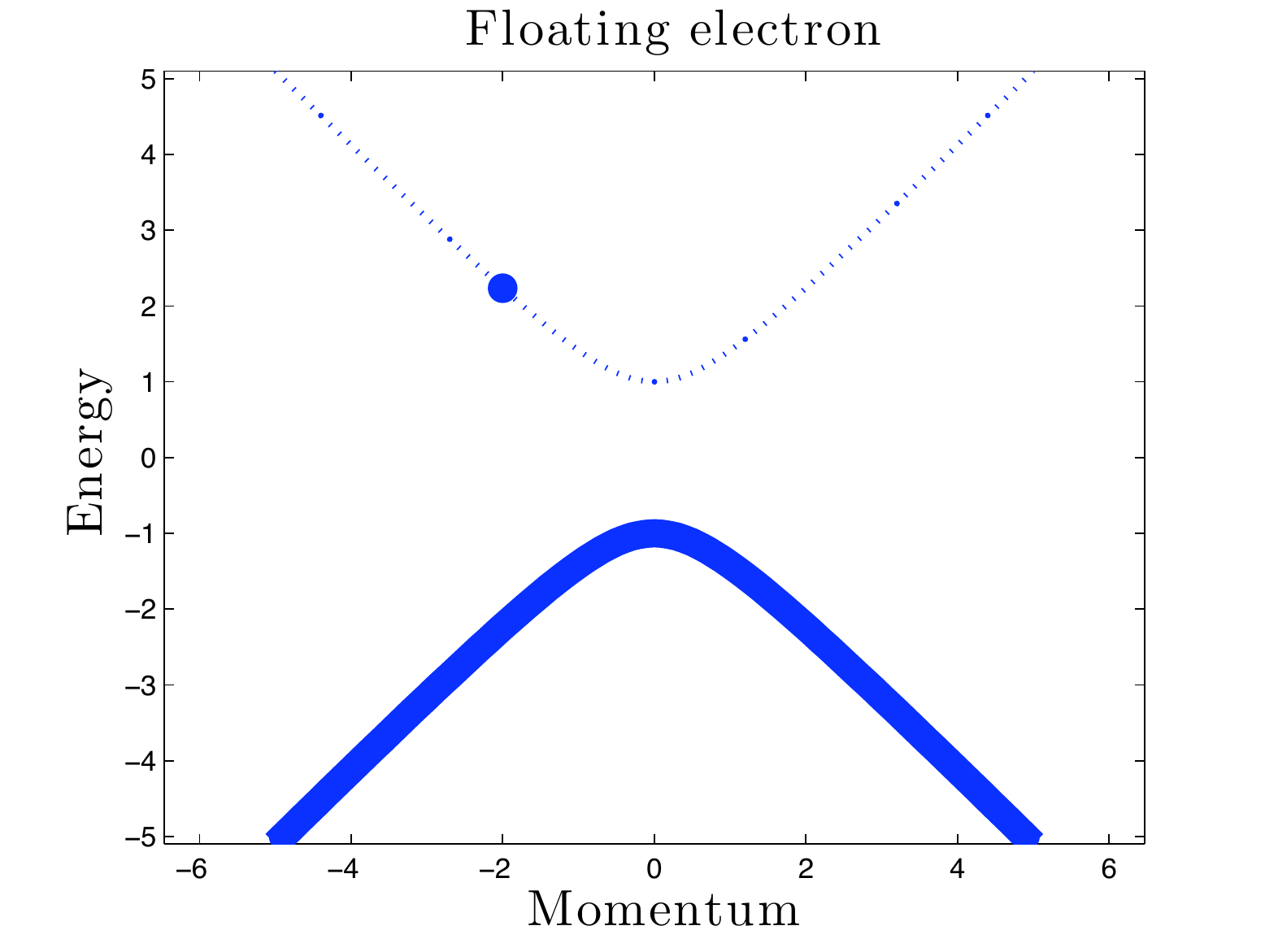}
\end{center}
\caption{\label{fig5}A literal single electron and a floating electron.}
\end{figure}

There are two ways to think about a state with a massive Dirac electron: it can be 
\begin{itemize}
\item either a Dirac electron on top of $|0_D\rangle$ (we refer to is as a literal single electron), 
\item or a Dirac electron on top of the sea $|G_D\rangle$ (we refer to it as a floating electron).
\end{itemize}
These two possibilities are illustrated in Fig. \ref{fig5}. 

\begin{figure}
\includegraphics[width=\textwidth]{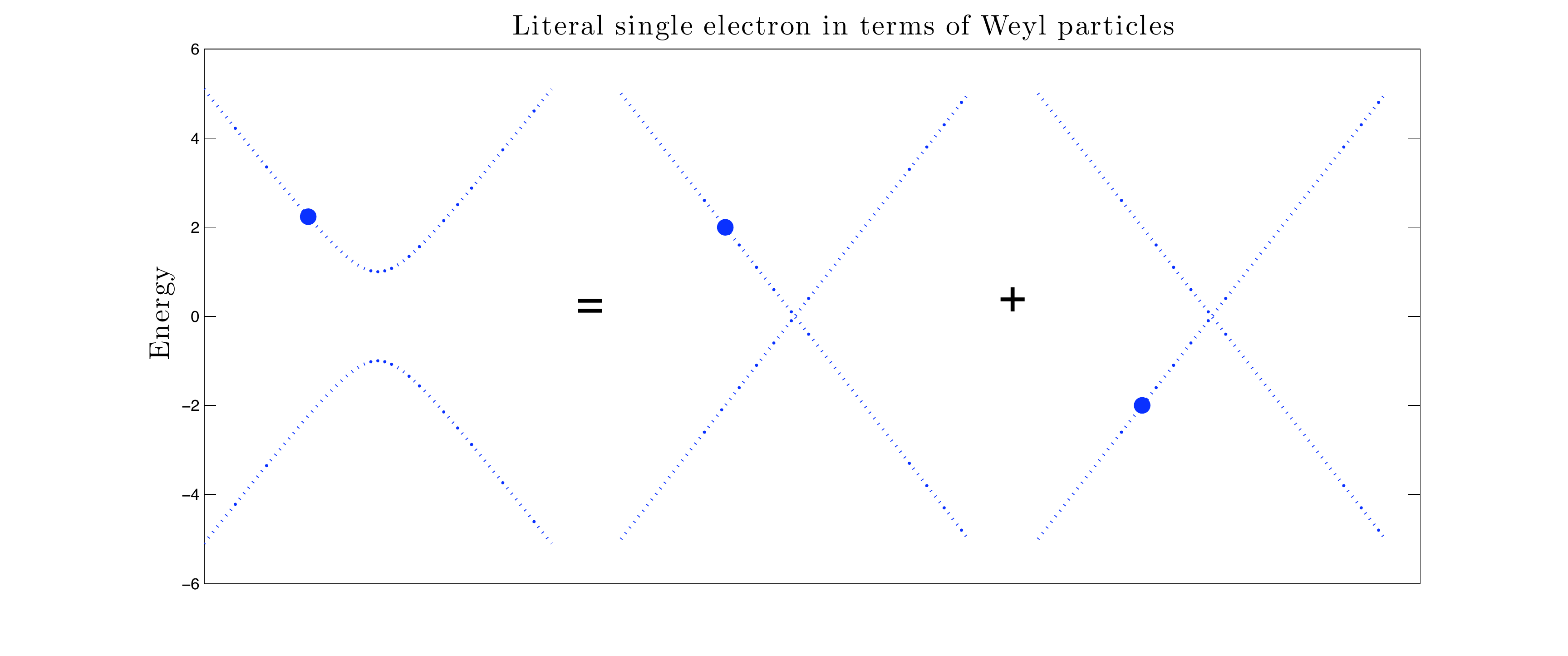}
\caption{\label{fig6}The representation of a literal single electron in terms of Weyl particles.}
\end{figure}

If we consider the first possibility, and if we look at Eq. (\ref{Hzeta}), we see that a state with a massive Dirac electron of momentum $\vec{p}$ and helicity $\chi$ 
can be written as a superposition  of states with positive and negative-energy Weyl particles:
\be
|e^-_\chi(\vec{p})\rangle=f_1(\vec{p})\hat{c}^\dagger_\chi(\vec{p})|0_D\rangle+f_2(\vec{p})\hat\zeta^\dagger_\chi(\vec{p})|0_D\rangle~.
\ee 
This is represented in Fig. \ref{fig6}.

\begin{figure}
\includegraphics[width=\textwidth]{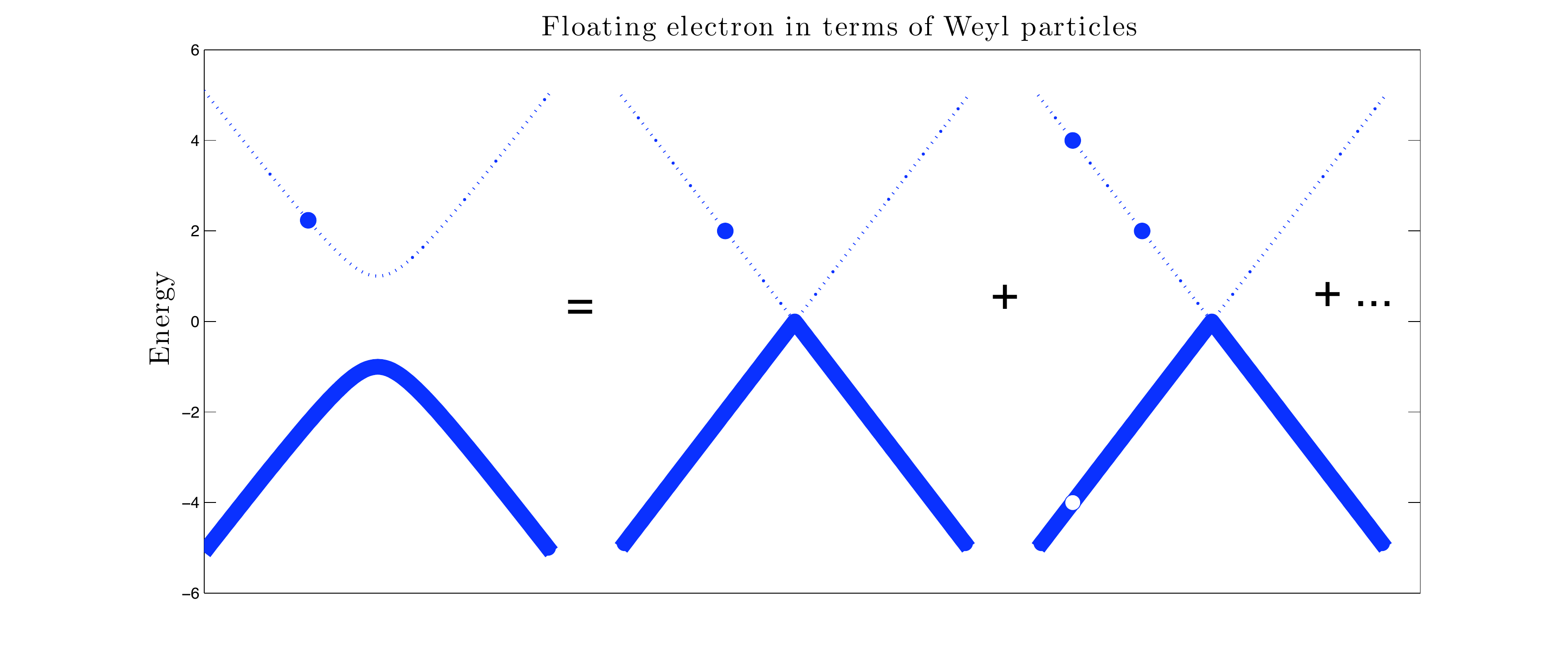}
\caption{\label{fig7}The representation of a floating electron in terms of Weyl particles.}
\end{figure}

Now we consider the second possibility. If we look at Eq. (\ref{Hd}), the state with a massive Dirac electron of momentum $\vec{p}$ and helicity $\chi$ 
can be written as a superposition:
\be
|e^-_\chi(\vec{p})\rangle=f_1(\vec{p})\hat{c}^\dagger_\chi(\vec{p})|G_D\rangle+f_2(\vec{p})\hat{d}_\chi(-\vec{p})|G_D\rangle~.
\ee
Therefore, thanks to the expression of $|G_D\rangle$ illustrated in Fig. (\ref{fig7a}), we find that the state of a massive Dirac electron is 
a state with a single Weyl fermion, plus a state with a single Weyl fermion and a pair, plus a state with a single Weyl fermion and two pairs etc. This is 
illustrated in Fig. \ref{fig7}.

Therefore viewing the Dirac electron as an oscillation between two Weyl particles only makes sense in the case of a literal single electron, for which 
\be
|e^-_\chi(\vec{p})\rangle=f_1(\vec{p})\hat{c}^\dagger_\chi(\vec{p})|0_W\rangle+f_2(\vec{p})\hat\zeta^\dagger_\chi(\vec{p})|0_W\rangle~.
\ee
 In order to get the coefficients $f_1(\vec{p})$ and $f_2(\vec{p})$, we diagonalize the following matrix:
\begin{align}
%\fl
\left(\begin{array}{cc} \langle 0_W|\hat{c}_\chi(\vec{p})\hat{H}\hat{c}^\dagger_\chi(\vec{p})|0_W\rangle & \langle 0_W|\hat\zeta_\chi(\vec{p})\hat{H}\hat{c}^\dagger_\chi(\vec{p})|0_W\rangle 
\\
\langle 0_W|\hat{c}_\chi(\vec{p})\hat{H}\hat\zeta_\chi^\dagger(\vec{p})|0_W\rangle &
 \langle 0_W|\hat\zeta_\chi(\vec{p})\hat{H}\hat\zeta_\chi^\dagger(\vec{p})|0_W\rangle\end{array}\right)=\left(\begin{array}{cc} p & m \\ m & -p \end{array}\right)~,\quad\quad
\end{align}
whose eigenvalues are $\sqrt{p^2+m^2}$ and $-\sqrt{p^2+m^2}$, where $p=|\vec{p}|$. 
The state describing the massive Dirac electron is then:
\begin{eqnarray}\label{zigzag}
|e^{-}_\chi(\vec{p})\rangle=\sqrt{\frac{E_p-p}{2E_p}}\left(\frac{m}{E_p-p}\hat{c}^\dagger_\chi(\vec{p})|0_W\rangle+\hat\zeta^\dagger_\chi(\vec{p})|0_W\rangle\right)
=\nonumber\\
\mathcal{N}_{c}(p)\hat{c}^\dagger_\chi(\vec{p})|0_W\rangle+\mathcal{N}_{\zeta}(p)\hat\zeta^\dagger_\chi(\vec{p})|0_W\rangle~,\quad\quad
\end{eqnarray}
where $E_p=\sqrt{p^2+m^2}$ and $\hat{H}|e^{-}_\chi(\vec{p})\rangle=E_p|e^{-}_\chi(\vec{p})\rangle$. This is the zig-zag electron: a superposition of states with positive and 
negative energy Weyl particles of the same helicity. In order to show explicitly how we can recover the Dirac equation, we consider a right-handed electron
\begin{align}
%\fl
|e^{-}_R(\vec{p}),t\rangle=e^{-iE_p t}\sqrt{\frac{E_p-p}{2E_p}}
\left(\frac{m}{E_p-p}\hat{c}^\dagger_R(\vec{p})|0_W\rangle+\hat\zeta^\dagger_R(\vec{p}|0_W\rangle)\right)\quad~,
\end{align}
and we define the quantities
\begin{align}
\Psi_{L,a}(t,\vec{x})=\langle 0_W|\hat\psi_{L,a}(\vec{x})|e^{-}_R(\vec{p}),t\rangle=
\sqrt{\frac{1}{(2\pi)^3}} u_{R,a}(\vec{p}) e^{-iE_p t}e^{i\vec{p}\cdot\vec{x}}\\
\Psi_{R,a}(t,\vec{x})=\langle 0_W|\hat\psi_{R,a}(\vec{x})|e^{-}_R(\vec{p}),t\rangle=
\sqrt{\frac{1}{(2\pi)^3}}\frac{m}{E_p-p}u_{R,a}(\vec{p})e^{-iE_p t}e^{i\vec{p}\cdot\vec{x}}~.
\end{align}
Then we can verify that the spinor obeys the Dirac equation:
\be
(i\gamma^\mu\partial_\mu-m)\left(\begin{array}{c}\Psi_L(t,\vec{x})\\ \Psi_R(t,\vec{x})\end{array}\right)=0~.
\ee
\section{The pilot-wave theory for the electron}
We now turn to the description of the trajectory of a single electron according to the pilot-wave theory. We contrast the zig-zag picture based on massless Weyl particles as advocated by Penrose (subsection A) with the conventional picture based directly on massive Dirac particles (subsection B). 
\subsection{The zig-zag picture}
Consider a positive-energy right-handed electron in a superposition of momentum eigenstates:
\be
|\Psi_t\rangle=\int d^3p \alpha(\vec{p}) e^{-iE_p t}|e^{-}_R(\vec{p})\rangle
\ee
Hence we have that:
\begin{align}
\Psi_{L,a}(t,\vec{x})=\langle 0_W|\hat\psi_{L,a}(\vec{x})|\Psi_t\rangle=
\sqrt{\frac{1}{(2\pi)^3}}\int d^3p \mathcal{N}_{c}(p)u_{R,a}(\vec{p})\alpha(\vec{p}) e^{-iE_p t}e^{i\vec{p}\cdot\vec{x}}\\
\Psi_{R,a}(t,\vec{x})=\langle 0_W|\hat\psi_{R,a}(\vec{x})|\Psi_t\rangle=
\sqrt{\frac{1}{(2\pi)^3}}\int d^3p \mathcal{N}_{\zeta}(p)u_{R,a}(\vec{p})\alpha(\vec{p}) e^{-iE_p t}e^{i\vec{p}\cdot\vec{x}}~.
\end{align}
If we want to construct the pilot-wave model corresponding to Penrose's zigzag picture of the electron, we can introduce mutually 
exclusive beables $\vec{x}_L(t)$ (zig) and $\vec{x}_R(t)$ (zag). In order to reproduce the predictions of the standard 
interpretation, the zig and the zag beables should be respectively distributed according to 
\be
\Psi^\dagger_L(t,\vec{x})\Psi_L(t,\vec{x})\textrm{ and }\Psi^\dagger_R(t,\vec{x})\Psi_R(t,\vec{x})
\ee
over an ensemble. One way to ensure this is by using the jump model introduced by D\"urr, Goldstein, Tumulka and Zangh\`{\i} \cite{dgtz04}, 
which can be thought of as a possible continuum generalization of the lattice stochastic model introduced by John Bell \cite{bell84}. In that case, 
$\vec{x}_L(t)$ would be guided by $\Psi_L(t,\vec{x})$:
\be
\vec{v}_L=-\frac{\Psi^\dagger_L\vec{\sigma}\Psi_L}{\Psi^\dagger_L\Psi_L}
\ee 
whereas $\vec{x}_R(t)$ would be guided by $\Psi_R(t,\vec{x})$:
\be
\vec{v}_R=\frac{\Psi^\dagger_R\vec{\sigma}\Psi_R}{\Psi^\dagger_R\Psi_R}~.
\ee 
The jiggling between the zig and zag is represented by a jump-rate; for instance, to switch from zag to zig, the rate is given by
\be
\sigma_t(\vec{y}|\vec{x})=2\frac{\left[\mathfrak{Im}\displaystyle\sum_{a,b}\Psi^\dagger_{L,a}(t,\vec{y}) H^{LR}_{I,a,b}(\vec{x},\vec{y})
\Psi_{R,b}(t,\vec{x})\right]^{+}}
{\Psi^\dagger_R(t,\vec{x})\Psi_R(t,\vec{x})}
\ee
where 
\be
H^{LR}_{I,a,b}(\vec{x},\vec{y})=\langle 0_W|\hat\psi_{L,a}(\vec{y})|\hat{H}_I|\hat\psi^\dagger_{R,b}(\vec{x})|0_W\rangle
\ee
and $[x]^+\equiv\max(x,0)$). The interaction Hamiltonian transition element is $\langle 0_W|\hat\psi_{L,a}(\vec{y})|\hat{H}_I|\hat\psi^\dagger_{R,b}(\vec{x})|0_W\rangle=m\delta_{ab}\delta^3(\vec{x}-\vec{y})$, so the jump-rate is given by:
\be
\sigma_t(\vec{y}|\vec{x})=2\frac{[\mathfrak{Im}(\Psi^\dagger_L(t,\vec{x})\Psi_R(t,\vec{x}))]^+}{\Psi^\dagger_R(t,\vec{x})\Psi_R(t,\vec{x})}m\delta^3(\vec{x}-\vec{y})~.
\ee

In the case of a momentum eigenstate ($\alpha(\vec{p})=\delta^3(\vec{p}-\vec{p}_0)$), the transition rate is equal to zero.

In the general case, we have that:
\begin{eqnarray}\label{psilpsir}
\Psi^\dagger_L(t,\vec{x})\Psi_R(t,\vec{x})=\frac{1}{(2\pi)^3}\int d^3p d^3 p' \alpha^*(\vec{p}\,') \alpha(\vec{p})\nonumber\\
e^{-i(E_{p}-E_{p'}) t}e^{i(\vec{p}-\vec{p}\,')\cdot\vec{x}}
\sqrt{\frac{E_{p'}-p'}{2 E_{p'}}} \frac{m}{\sqrt{2 E_p(E_p-p)}} \nonumber\\\left(\begin{array}{cc} 1 & 0\end{array}\right)(p'+\vec{\sigma}\cdot\vec{p}\,')(p+\vec{\sigma}\cdot\vec{p})\left(\begin{array}{c} 1 \\ 0\end{array}\right) \mathcal{N}_{\vec{p}\,'}\mathcal{N}_{\vec{p}}~,~~
\end{eqnarray}
where $\mathcal{N}_{\vec{p}}$ is the normalization factor of the helicity eigenstate. The helicity factor becomes
\begin{align}
\left(\begin{array}{cc} 1 & 0\end{array}\right)(p'+\vec{\sigma}\cdot\vec{p}\,')(p+\vec{\sigma}\cdot\vec{p})\left(\begin{array}{cc} 1 \\ 0\end{array}\right) \mathcal{N}_{\vec{p}\,'}\mathcal{N}_{\vec{p}}=
\frac{pp'+p'p_z+pp'_z+\vec{p}\cdot\vec{p}'}{\sqrt{2p(p+p_z)}\sqrt{2p'(p'+p'_z)}}~.
\end{align}
If the electron is moving along the z-direction, the last expression can be further simplified too:
\be
\frac{1}{2}\sqrt{\frac{(p+p_z)(p'+p'_z)}{p p'}}~,
\ee
where $p=|p_z|$ and $p'=|p'_z|$. For motion in the positive z-direction, this is equal to 1. If we further assume that the particle momentum is in the non-relativistic regime, 
we have that:
\begin{align}
\frac{m}{\sqrt{2 E_p(E_p-p)}}\sqrt{\frac{E_{p'}-p'}{2 E_{p'}}}\simeq\frac{1}{2}(1+\frac{1}{2m}(p-p'))=\frac{1}{2}(1+\frac{1}{2m}(p_z-p'_z))~.~~
\end{align}
Then, if we define
\be
\phi(t,\vec{x})=\sqrt{\frac{1}{(2\pi)^3}}\int d^3p \alpha(\vec{p}) e^{-iE_p t}e^{i\vec{p}\cdot\vec{x}}~,
\ee
we have, for motion in the positive z-direction, that 
\begin{eqnarray}
\Psi^\dagger_L(t,\vec{x})\Psi_R(t,\vec{x})\simeq\frac{1}{2}\phi^\dagger(t,\vec{x})\phi(t,\vec{x})-\frac{i}{4m}\phi^\dagger(t,\vec{x})\partial_z\phi(t,\vec{x})\nonumber\\ \rightarrow
\mathfrak{Im}(\Psi^\dagger_L(t,\vec{x})\Psi_R(t,\vec{x}))=-\frac{1}{8m} \partial_z|\phi(t,\vec{x})|^2~.~\quad\quad
\end{eqnarray}
We still have to evaluate the quantity $\Psi^\dagger_R(t,\vec{x})\Psi_R(t,\vec{x})$ in the same limit (it appears in the denominator of the jump-rate). Along the same lines, 
we obtain that:
\be
\Psi^\dagger_L(t,\vec{x})\Psi_R(t,\vec{x})\simeq \frac{1}{2}\phi^\dagger(t,\vec{x})\phi(t,\vec{x})~.
\ee 
Thus finally, for the jump-rate in the non-relativistic limit, we obtain
\begin{align}
\sigma_t(\vec{y}|\vec{x})=\frac{1}{4}\frac{[\partial_z|\phi(t,\vec{x})|^2]^-}{|\phi(t,\vec{x})|^2}\delta^3(\vec{x}-\vec{y})=
\frac{1}{4}[\partial_z\ln{|\phi(t,\vec{x})|^2}]^{-}\delta^3(\vec{x}-\vec{y})~.
\end{align}
As an example, we consider a Gaussian wave-packet localized a the origin
\be
|\phi(t,\vec{x})|^2=\mathcal{N}_t e^{\frac{-|\vec{x}|^2}{2\sigma_t^2}}~.
\ee
The jump-rate is then given by
\be
\sigma_t(\vec{y}|\vec{x})=\frac{[z]^+}{4\sigma_t^2}\delta^3(\vec{x}-\vec{y})~.
\ee
That is, if and only if the electron is in the right-hand side of the packet, there is a probability to jump, and the probability gets smaller as the packet spreads.
\subsection{The Dirac picture}
We now contrast the above with the usual deterministic pilot-wave theory for the electron, by tracing out over the indices $L$ and $R$. 
We note that in the conventional Dirac picture, the mass is not interpreted as an interaction. 

We define the standard probability density
\be
\rho_S(t,\vec{x})=\sum_{\chi}\Psi^\dagger_\chi(t,\vec{x})\Psi_\chi(t,\vec{x})~,
\ee
where
\be
\Psi_{\chi,a}(t,\vec{x})=\langle 0_W|\hat\psi_{\chi,a}(\vec{x})|\Psi(t)\rangle~,
\ee
 and we take its time-derivative, in order to find: 
\begin{eqnarray}
\frac{\partial\rho_S(t,\vec{x})}{\partial t}-\vec{\nabla}\cdot\Psi^\dagger_L\vec{\sigma}\Psi_L+\vec{\nabla}\cdot\Psi^\dagger_R\vec{\sigma}\Psi_R=0~.
\end{eqnarray}
Both currents are light-like but their difference won't be. Nevertheless, we can define the pilot-wave theory by defining the electron velocity as
\be
\vec{v}=\frac{\Psi^\dagger_R\vec{\sigma}\Psi_R-\Psi^\dagger_L\vec{\sigma}\Psi_L}{\Psi^\dagger_L\Psi_L+\Psi^\dagger_R\Psi_R}~.
\ee
The pilot-wave theory is equivalent to the one presented at Subsection B of Section 3; the conceptual difference is that in this Subsection, we have shown 
how it can be derived from a second quantized theory.
\subsection{Example}
In order to illustrate the differences between the trajectories predicted by the two theories, we consider the following state:
\be
|\Psi_t\rangle=\frac{1}{\sqrt{3}}(|e^-_R(\vec{p}_1),t\rangle+e^{i4}|e^-_R(\vec{p}_2),t\rangle+e^{i9}|e^-_R(\vec{p}_3),t\rangle)~,
\ee
where $\vec{p}_1=(1,0,1)$, $\vec{p}_2=(-1,-2,-1)$ and $\vec{p}_3=(1,-1,1)$. The electron has mass $m=10$, its initial position is 
$(0,1,0)$ and the trajectories run from $t=0$ to $t=50$. The blue trajectory corresponds to 
the zig-zag picture while the red one corresponds to the Dirac picture. The points where the blue trajectory changes direction discontinuously correspond to jumps from zig to zag motion or vice-versa. 
Phases of intensive jiggling can be interrupted by long periods of smooth motion. This is most likely due to this scenario: the beable is in the zig (resp. zag) 
motion and enters a region where the imaginary part of $\Psi^\dagger_R\Psi_L$ (resp. $\Psi^\dagger_L\Psi_R$) is negative, hence 
jumps are forbidden, and the electron remains in zig (resp. zag) motion until the region is traversed.
\begin{figure}
\includegraphics[width=\textwidth]{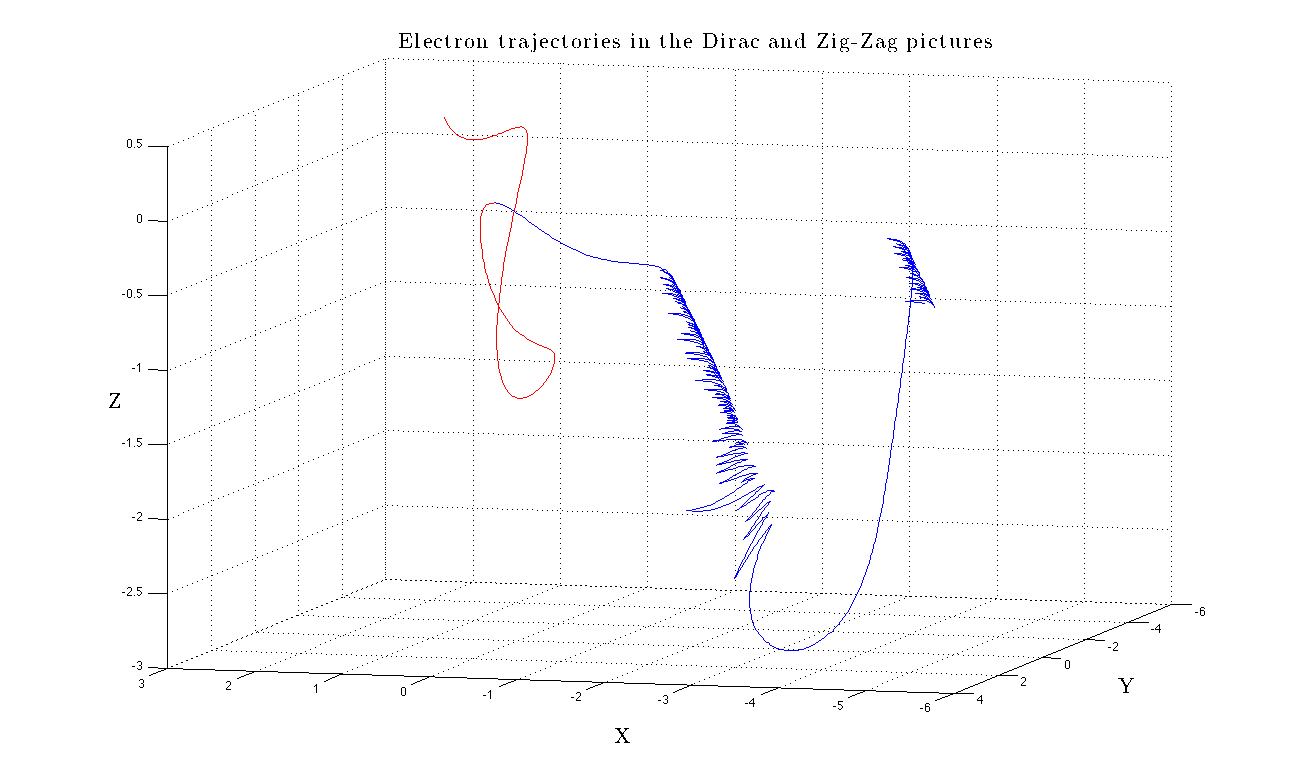}
\caption{\label{fig8}The red trajectory corresponds to the conventional massive Dirac picture. 
The blue trajectory corresponds to the zig-zag picture. The points where the blue trajectory changes direction discontinuously 
correspond to jumps from zig to zag motion or vice-versa. 
For more details, see text.}
\end{figure}
\section{Weyl seas and pilot waves}
Now we address the more general question: assuming that the beables should be attributed to massless fermions, what is the most compelling 
pilot-wave model for the standard model of particle physics? 
The zig-zag picture has shown us that negative energies should be considered seriously: a positive-energy electron is a superposition of positive and negative 
energy Weyl particles. A negative-energy picture also has the advantage that the total number of fermions (total 
number of positive and negative energy fermions) is conserved, which makes it possible to have a deterministic pilot-wave model.

The question is whether we should adopt a Weyl ontology or a Dirac ontology. In the Weyl ontology, the beables are attributed to Weyl particles 
and they always move luminally, like in the zigzag picture, but determinism needs to be sacrified. 
On the other hand, in the massless Dirac ontology, we trace out over $R$ and $L$ labels and we attribute beables to massless Dirac particles: 
the theory is deterministic but beables do not move luminally anymore. 

In considering this issue it should be borne in mind that if we consider two Weyl fermions, their individual motions are not luminal anymore. 
Take for example two Weyl fermions in an $R$-state, that is 
\be
|\Psi_t\rangle=\sum_{a_1,a_2}\Psi_{a_1,a_2}(t,\vec{x}_1,\vec{x}_2)\hat{\psi}_{R,a_1}(\vec{x}_1)\hat{\psi}_{R,a_2}(\vec{x}_2)|0_W\rangle~.
\ee
Then the velocities are given by
\begin{eqnarray}
\vec{v}_1=\frac{\Psi^*_{a_1 a_2}(t,\vec{x}_1,\vec{x}_2)\vec{\sigma}_{a_1 a}\Psi_{a a_2}(t,\vec{x}_1,\vec{x}_2)}
{\Psi^*_{b_1 b_2}(t,\vec{x}_1,\vec{x}_2)\Psi_{b_1 b_2}(t,\vec{x}_1,\vec{x}_2)}~\label{eq66}\\
\vec{v}_2=\frac{\Psi^*_{a_1 a_2}(t,\vec{x}_1,\vec{x}_2)\vec{\sigma}_{a_2 a}\Psi_{a_1 a}(t,\vec{x}_1,\vec{x}_2)}
{\Psi^*_{b_1 b_2}(t,\vec{x}_1,\vec{x}_2)\Psi_{b_1 b_2}(t,\vec{x}_1,\vec{x}_2)}~.\label{eq67}
\end{eqnarray}
If we introduce $\Psi_{ij}=R_{ij}e^{i\theta_{ij}}$, use the definition of the $\sigma$-matrices, we find that 
\begin{eqnarray}
\vec{v}_1\cdot\vec{v}_1 =1-\frac{4}{\rho^2}(R^2_{11}R^2_{22}+R^2_{21}R^2_{12}\nonumber\\
-2\cos(\theta_{11}+\theta_{22}-\theta_{12}-\theta_{21})R_{11}R_{22}R_{12}R_{21})~,
\end{eqnarray}
where $\rho_\Psi(t,\vec{x}_1,\vec{x}_2)=\Psi^*_{b_1 b_2}(t,\vec{x}_1,\vec{x}_2)\Psi_{b_1 b_2}(t,\vec{x}_1,\vec{x}_2)$. 
Clearly the velocity is in general not luminal anymore when we consider two particles.
If the particles are not entangled then they will individually be luminal, as is clear from Eq. (\ref{eq66}) and Eq. (\ref{eq67}). 
But if they are identical particles then their wavefunction should be symmetrized so that they will be entangled. 
\section{Conclusion}
In conclusion, we have:
\begin{itemize}
\item pointed out the importance of bare particle ontologies, in particular of massless fermions in the case of the standard model for particle physics, 
and why this might be relevant for Valentini's hypothesis relating to quantum non-equilibrium, 
\item shown that the zig-zag electron is a superposition of positive and negative-energy Weyl particles of the same helicity, and not a superposition of Weyl particles 
of opposite helicity as suggested by Penrose,
\item shown, in the pilot-wave model corresponding to the zigzag picture, that a single electron can in principle move luminally at all times, 
\item shown that the Dirac ontology has the advantage of determinism over the Weyl ontology, 
while the latter's advantage of luminality vanishes as soon as one considers more than one particle.
\end{itemize}
The present work can also be relevant for pilot-wave models for supersymmetric quantum field theories because Majorana or Weyl spinors 
are usually associated to bosonic vector fields.

Also, even if it does not seem to work in the present case, the idea of luminal beables is a very compelling one 
(new insight on quantum non-locality and Lorentz invariance?), and it might be worthwhile to look further into it in different theories (for example light-front quantized 
quantum field theory \cite{heinzl}) or other hidden-variable models, that depart from the standard de Broglie-Bohm construction 
(for instance psi-epistemic theories \cite{montina,westman}).

Another topic for future work is to address the issues which are relevant to the construction of a pilot-wave theory of massless 
fermions before spontaneous symmetry breaking (fermions on expanding space and tachyonic quantum field theories).
\section*{Acknowledgments}
S. C. thanks in particular Ward Struyve for stimulating discussions. He also thanks Hans Westman for discussions in an early stage, 
Thomas Durt and Antony Valentini for comments on an early draft, and acknowledges financial support from a 
Perimeter Institute Australia Foundations postdoctoral research fellowship. 
This work was also supported by the Australian Research Council Discovery Project DP0880657, ``Decoherence, Time-Asymmetry and the Bohmian View on the 
Quantum World''.
Part of this work was done at Perimeter Institute. Research at Perimeter Institute is funded by the Government of Canada through Industry Canada and by the 
Province of Ontario through the Ministry of Research and Innovation.
% Create the reference section using BibTeX:
%\section*{References}


\begin{thebibliography}{10}

\bibitem{perez}
Alejandro Perez, Hanno Sahlmann, and Daniel Sudarsky.
\newblock On the quantum origin of the seeds of cosmic structure.
\newblock {\em Class. Quantum Grav.}, 23(7):2317--2354, March 2006.

\bibitem{valentini08}
Antony Valentini.
\newblock Inflationary cosmology as a probe of primordial quantum mechanics.
\newblock {\em Phys. Rev. D}, 82(6):063513, Sep 2010.

\bibitem{mukhanov}
V.\ Mukhanov.
\newblock {\em Physical Foundations of Cosmology}.
\newblock Cambridge University Press, 2005.

\bibitem{debroglie28}
{L.\ de Broglie, in ``Electrons et Photons: Rapports et Discussions du
  Cinqui\`eme Conseil de Physique'', ed.\ J.\ Bordet, Gauthier-Villars, Paris,
  105 (1928), English translation: G.\ Bacciagaluppi and A.\ Valentini,
  ``Quantum Theory at the Crossroads: Reconsidering the 1927 Solvay
  Conference'', Cambridge University Press (2009), also quant-ph/0609184.}

\bibitem{bohm521}
{D.\ Bohm, {\em Phys.\ Rev.}\ {\bf 85}, 166 (1952).}

\bibitem{bohm522}
{D.\ Bohm, {\em Phys.\ Rev.}\ {\bf 85}, 180 (1952).}

\bibitem{peterpintoneto}
Patrick Peter and Nelson Pinto-Neto.
\newblock Cosmology without inflation.
\newblock {\em Phys. Rev. D}, 78(6):063506, Sep 2008.

\bibitem{cost07}
{S.\ Colin and W.\ Struyve, {\em J.\ Phys.\ A} {\bf 40}, 7309--7341 (2007) and
  quant-ph/0701085.}

\bibitem{dgtz04}
{D.\ D\"urr, S.\ Goldstein, R.\ Tumulka and N.\ Zangh\`\i, {\em Phys.\ Rev.\
  Lett.}\ {\bf 93}, 090402 (2004) and quant-ph/0303156.}

\bibitem{feinberg}
G.~Feinberg.
\newblock Possibility of faster-than-light particles.
\newblock {\em Phys. Rev.}, 159(5):1089--1105, Jul 1967.

\bibitem{stwe}
{H. Westman and W. Struyve, {\em Proc. R. Soc. A} {\bf 463}, 3115--3129 (2007).}

\bibitem{penroseroad}
R.~Penrose.
\newblock {\em {The Road to Reality}}.
\newblock Alfred A. Knopf, 2005.

\bibitem{valentini-phd}
{A.\ Valentini, ``On the Pilot-Wave Theory of Classical, Quantum and Subquantum
  Physics", PhD.\ Thesis, International School for Advanced Studies, Trieste
  (1992), online \url{http://www.sissa.it/ap/PhD/Theses/valentini.pdf}.}

\bibitem{holland}
P.~R. Holland.
\newblock {\em {The Quantum Theory of Motion}}.
\newblock Cambridge University Press, 1993.

\bibitem{wiseman2007}
{H. M. Wiseman, {\em New J. Phys.} {\bf 9}, 165 (2007).}

\bibitem{durr92}
{D.\ D\"urr, S.\ Goldstein and N.\ Zangh\`\i, {\em J.\ Stat.\ Phys.}\ {\bf 67},
  843 (1992) and quant-ph/0308039.}

\bibitem{valentini042}
{A.\ Valentini and H.\ Westman, {\em Proc.\ R.\ Soc.\ A} {\bf 461}, 253 (2005)
  and quant-ph/0403034.}

\bibitem{cost10}
{S.\ Colin and W.\ Struyve, {\em New\ J.\ Phys.} {\bf 12}, 043008 (2010).}

\bibitem{struyve}
{W. Struyve, {\em Rep. Prog. Phys.} {\bf 73}, 106001 (2010).}

\bibitem{bell84}
{J.S.\ Bell, CERN preprint CERN-TH. 4035/84 (1984), reprinted in J.S.\ Bell,
  ``Speakable and unspeakable in quantum mechanics", Cambridge University
  Press, Cambridge (1987).}

\bibitem{colin_paon}
{S. Colin, Particle ontologies in pilot-wave quantum field theories.}

\bibitem{toruva}
{Towler, M. D. and Russell, N. J. and Valentini, A., Timescales for dynamical
  relaxation to the Born rule, 1103.1589}.

\bibitem{efthymiopoulos}
C.~Efthymiopoulos, C.~Kalapotharakos, and G.~Contopoulos.
\newblock Origin of chaos near critical points of quantum flow.
\newblock {\em Phys. Rev. E}, 79(3):036203, Mar 2009.

\bibitem{nature}
{R. Gerritsma et al., {\em Nature} {\bf 463}, 68--72 (2010).}

\bibitem{tatu}
{D. V. Tausk and R. Tumulka, {\em J. Math. Phys.} {\bf 51}, 122306 (2010).}

\bibitem{heinzl}
T.~Heinzl.
\newblock {Light-cone quantization: foundations and applications}.
\newblock In {\em {Lect.Notes Phys. 572 (2001) 55-142}}, 2001.

\bibitem{montina}
A.~Montina.
\newblock Exponential complexity and ontological theories of quantum mechanics.
\newblock {\em Phys. Rev. A}, 77(2):022104, Feb 2008.

\bibitem{westman}
{H.\ Westman, ``A candidate of a psi-epistemic theory'', online talk
  \url{http://pirsa.org/08070033/}.}

\end{thebibliography}
\end{document}